\documentclass[aps,prl,twocolumn,superscriptaddress,showpacs,longbibliography]{revtex4-2}
\usepackage{graphicx} 
\usepackage{amsmath,amssymb,amsfonts,float,graphics,epsfig,epstopdf,color,verbatim,tabularx,bm,multirow,appendix}
\usepackage{xcolor}
\usepackage{dsfont}
\usepackage{textcomp}
\usepackage{bm}

\usepackage{graphicx}%
\usepackage{xcolor}
\usepackage{dcolumn}
\usepackage{bm}
\usepackage{xspace}
\usepackage{time}
\usepackage{booktabs}
\usepackage{multirow}
\usepackage{makecell}

\pdfoutput=1
\usepackage{color}
\definecolor{LinkColor}{rgb}{0.256,0.439,0.588}
\usepackage{hyperref}
\hypersetup{
colorlinks=true,
citecolor=LinkColor,
linkcolor=LinkColor,
urlcolor=LinkColor
}
\usepackage{cleveref}
\Crefname{equation}{Eq.}{Eqs.}
\Crefname{figure}{Fig.}{Figs.}

\graphicspath{{./figures/}}

\usepackage{stackengine}

\begin{document}
\title{Non-Abelian braiding in Abelian Fractional Quantum Hall Phases from realistic interactions}
\author{Ha Quang Trung}
\affiliation{Division of Physics and Applied Physics, Nanyang Technological University, Singapore 637371, Singapore}

\author{Bo Yang}
\email{yang.bo@ntu.edu.sg}
\affiliation{Division of Physics and Applied Physics, Nanyang Technological University, Singapore 637371, Singapore}

\date{\today}

\begin{abstract}
We propose a method of realizing non-Abelian braiding of fractionalized quasiholes in the Laughlin fractional quantum Hall phase at $\nu=1/3$ with realistic two-body interactions within the lowest Landau level. It is numerically shown that low-lying gapped excitations near $\nu=1/3$ are contained almost entirely within the null space of the three-body Moore-Read model Hamiltonian. They are thus quantum fluids of non-Abelian quasiholes that are in principle physically accessible. In particular, Laughlin ground state can be described as a fluid of ``$\psi$-type" quasiholes formed by binding a magnetic flux with a Majorana fermion (MF), and the Laughlin quasiholes are described by the ``$1$-type'' quasiholes, which are magnetic fluxes without a MF attached. Within the Laughlin phase, Laughlin quasiholes can be locally fractionalized into non-Abelian quasiholes, when the strong attraction between them is overcome by properly designed one-body electrostatic trapping potentials. Extensive numerics with proper finite-size scaling corroborate this physical picture, and our study points to the possibility of realizing non-Abelian braiding within an Abelian topological phase in experiment without the need for fine-tuning realistic electron-electron interaction.
\end{abstract}

\maketitle

\textit{Introduction} -- 
One of the most interesting aspects of strongly correlated two-dimensional topological materials is the emergence of anyonic excitations: collective particle-like excitations with non-trivial exchange statistics \cite{Leinaas1977,wilczek1990fractional,simon2023topological}. Anyons are broadly divided into two categories: Abelian and non-Abelian. Abelian anyons are generalization of bosons and fermions -- exchanging two particles results in a scalar phase that interpolates between 0 and $\pi$ \cite{Leinaas1977}, and non-Abelian anyons further generalize this statistics by promoting this scalar phase into a tensor quantity \cite{simon2023topological}. This requires at least a two-fold degeneracy in the many-particle wavefunctions even when the position of every anyon is fixed. Thus, non-Abelian braiding necessarily requires non-trivial degeneracy of quantum states of a collection of anyons \emph{in addition} to the degeneracy from their different real space positions.

There have been many proposals for realizing non-Abelian anyons in condensed matter systems and designing topological quantum computing (TQC) protocols on such systems \cite{Nayak2008,Freedman2000,sarma2015majorana,kitaev2003fault,das2005topologically}. The fractional quantum Hall (FQH) systems is one of the earliest platforms realized in experiments where anyons have been predicted \cite{tsui1982two,yoshioka1984ground,arovas1984fractional,moore1991nonabelions,nayak19962n,Read1996,feldman2021}. Other systems capable of hosting non-Abelian anyons, in particular Majorana fermions, include topological superconductors (including superconducting wires) \cite{kitaev2001unpaired,alicea2012new,hu2015majorana,franz2010race} and spin chain \cite{kitaev2003fault,kitaev2006anyons,tsintzis2024majorana,li2014probing}. Ground-breaking experimental progresses on these platforms have only been reported recently \cite{google2023non,aghaee2023inas,microsoft2025interferometric,aasen2025roadmap}, but not without controversies \cite{frolov2021quantum,rini2025microsoft}. In general, the key challenge in experiment is that while anyon-based TQC itself is intrinsically noise-resistant, establishing a topological order to host these anyons in the first place requires very pristine conditions \cite{aghaee2023inas}.

Studies done on FQH systems face similar challenges in establishing strong experimental evidences for non-Abelian anyons. Current research focuses on the most promising non-Abelian state currently experimentally realized at the filling factor $\nu=5/2$ \cite{willett1987observation,pan1999exact,willett2023interference}. This FQH plateau is believed to be described by the Moore-Read (Pfaffian) wavefunction \cite{moore1991nonabelions} or its particle-hole conjugate \cite{levin2007particle,wang2018topological, ma2022fractional}, and with controversies \cite{wang2018topological,asasi2020partial,hein2023thermal,manna2024full}. In experiments, the fractional charge of the quasiholes of the $\nu=5/2$ FQH plateau has been measured, which supports either of the model states in the Pfaffian family \cite{dolev2008observation,venkatachalam2011local,kim2026aharonov}. However, while all of these model states are non-Abelian, experimental evidences of their non-Abelian statistics are few and inconclusive \cite{willett2023interference}. Another popular non-Abelian state is the so-called Fibonacci phase from the Read-Rezayi series, a suitable candidate for universal quantum computing \cite{read1999beyond,Nayak2008}. While it is believed to be a model state describing the filling factor $\nu=12/5$, to date there has been no experimental evidence supporting the non-Abelian properties at this filling factor \cite{rezayi2009non,wojs2009transition,zhu2015fractional,Wu2017a}. Nevertheless the existence of these exotic topological phases means the FQH system remains a platform with many unexplored physical phenomena that promise transformative technologies.

\begin{figure}
\begin{center}
\includegraphics[width=\linewidth]{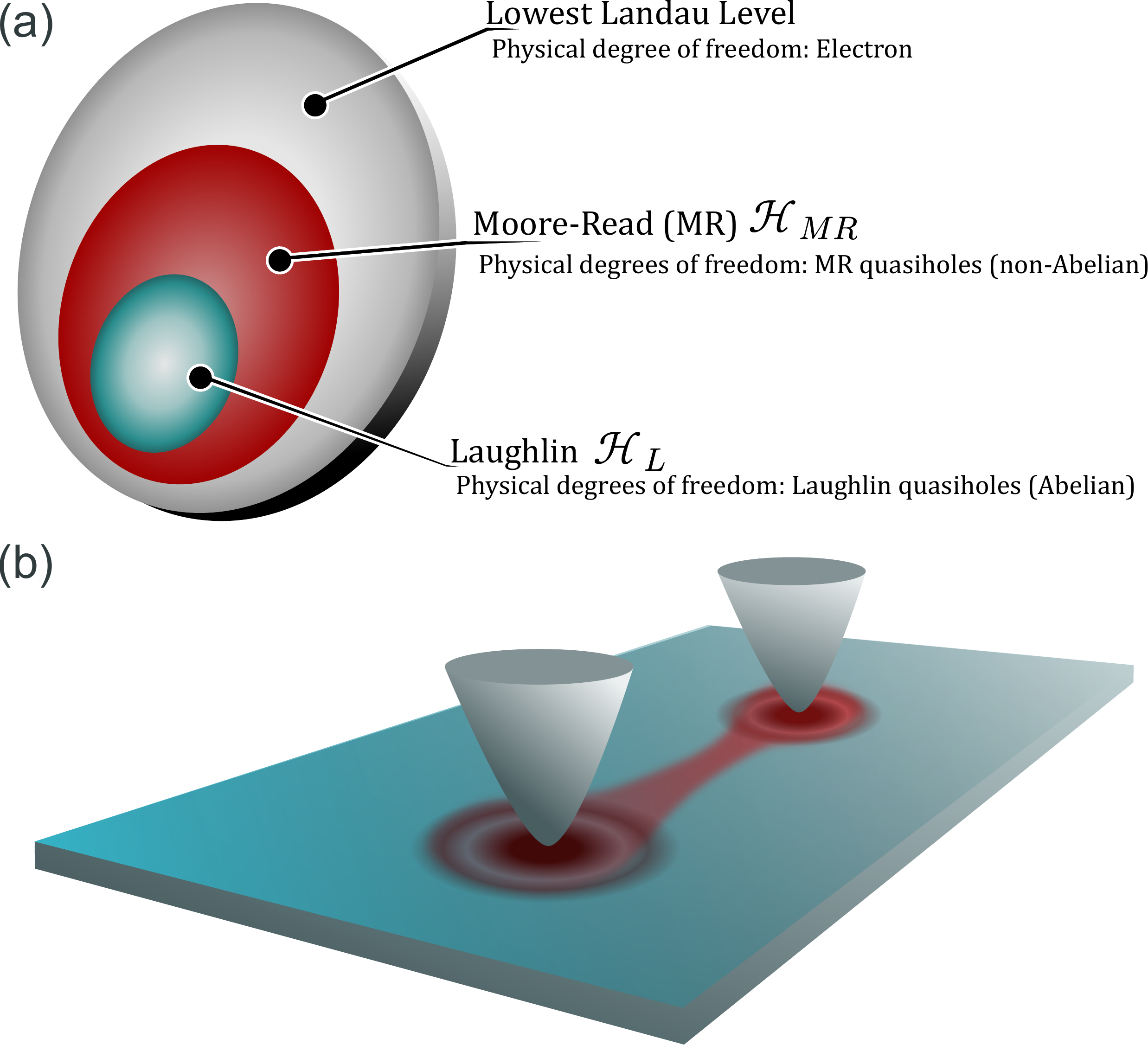}
\caption{(a) The conformal Hilbert space hierarchy showing the relationship between the lowest Landau level, the Moore-Read subspace $\mathcal H_{\text{MR}}$, and the Laughlin subspace $\mathcal H_{\text{L}}$. (b) A schematic diagram showing fractionalization of an Abelian quasihole by electrostatic potential induced by two wide STM tips. A region of density oscillation is formed between two half-quasiholes.}
\label{fig:Illustration}
\end{center}
\end{figure}

Motivated by this, we investigate the possibility of realizing non-Abelian statistics within an Abelian FQH phase, which is much more stable in realistic systems. Our study is informed by recent progresses in the theory of conformal Hilbert space (CHS) \cite{wang2022analytic,yang2022anyons}, which have uncovered a fundamental relationship between different FQH phases (and the corresponding anyonic excitations), both Abelian and non-Abelian.
An illustrative example is the easily realizable Laughlin phase at filling factor $\nu=1/3$: its model Hamiltonian defines a null space of many-body states within a single Landau level, physically spanned by all the zero energy ground states and quasiholes (the Abelian anyons) \cite{laughlin1983anomalous,cage2012quantum,yoshioka1984ground,haldane1983fractional}. This is a subspace with emergent conformal symmetry, thus denoted as the Laughlin CHS denoted by $\mathcal H_{\text{L}}$. One can prove $\mathcal H_{\text{L}}$ is the proper subspace of another CHS denoted by $\mathcal H_{\text{MR}}$, which is spanned by the ground state and quasiholes (the non-Abelian anyons) of the Moore-Read model Hamiltonian~\cite{wang2022analytic}. 

The fact that $\mathcal H_{\text{L}}\subseteq\mathcal H_{\text{MR}}$ has an important implication for the low-energy states of the Laughlin phase: for all relevant physics we can understand them as quantum fluids of (bound) non-Abelian Moore-Read anyons (see \Cref{fig:Illustration}), which we can show to be weakly interacting. In particular, the gapped excitations can contain unbound non-Abelian anyons that may be physically accessible \cite{yang2022anyons,trung2021fractionalization,wang2026microscopic}. The possibility of the fractionalization of Laughlin quasiholes near the nematic FQH phase was also studied in detail in Ref. \cite{trung2021fractionalization}, which suggests the existence of charge-$e/6$ quasiholes (also appearing in \cite{balram2020z}) at $\nu=1/3$ filling. This physics is well captured within $\mathcal H_{\text{MR}}$, and thus one may design experimental schemes to fractionalize the Laughlin quasiholes and exploit their non-Abelian properties.

Instead of the nematic FQHE with soft neutral excitations (requiring difficult fine-tuning of electron-electron interaction), in this work we focus on the easily realizable fully gapped Laughlin state at $\nu=1/3$, and show that while all the physics below the incompressibility gap is universal and Abelian, non-Abelian statistics can be experimentally accessed from the low-lying \emph{gapped} states. Very interestingly, it can be established that the ground state, quasihole states and low-lying gapped excitations of the Laughlin phase are quantum fluids of Moore-Read (MR) quasiholes naturally emerging from \emph{realistic two-body interaction}. We can thus show that the non-Abelian quasiholes can be realized by the fractionalization of the Abelian quasiholes in the real space, using local one-body trapping potentials. Conceptually, this scheme describes an intriguing localized phase transition (i.e. level crossing) deep within the correlated topological fluid involving only a few local quasiholes; experimentally it leads to a rather novel approach for realizing non-Abelian braiding in two-dimensional topological materials.

\begin{figure*}
\begin{center}
\includegraphics[width=\linewidth]{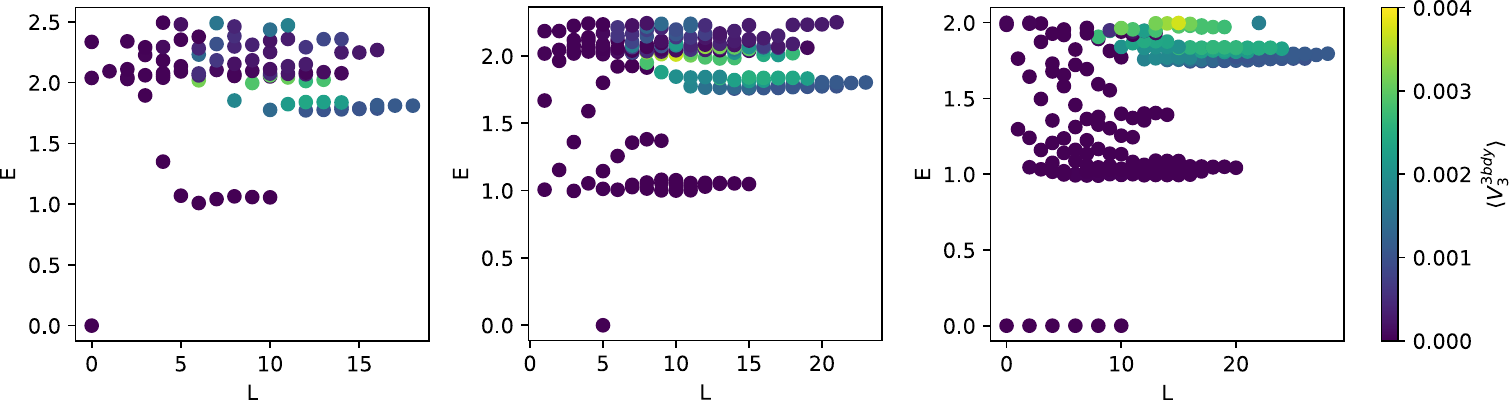}
\caption{The spectrum of $\hat V_1^{2bdy}$ on the sphere with 10 electrons and 28, 29, and 30 orbitals (Laughlin ground state, one-quasihole state, and two-quasihole state), respectively. The lowest 1000 eigenvalues are shown in each plot. The color bar shows variational energy of each eigenstate with respect to the Moore-Read model Hamiltonian, $\hat V_3^{3bdy}$. All numerical results in this paper are given in the unit $\hbar=c=e=B=1$.}
\label{fig:3-body energies}
\end{center}
\end{figure*}

\textit{Laughlin state as a quantum fluid of non-Abelions} -- We start by showing that not only the gapless quasiholes, but also the low-lying gapped excitations of the Laughlin phase are quantum fluids of non-Abelian anyons of the MR state. While for quasiholes this is analytically proven, for gapped excitations we can show numerically using the spherical geometry \cite{haldane1983fractional}, where the states in the Laughlin CHS $\mathcal H_{\text{L}}$ consist of $N_o$ orbitals and $N_e$ electron satisfying $N_o=3N_e-2+N_{q}$, where $N_q$ is the number of Laughlin quasiholes. The CHS $\mathcal H_{\text{L}}$ and $\mathcal H_{\text{MR}}$ are the nullspaces of the two-body pseudopotential $\hat V_1^{2bdy}$ and the three-body pseudopotential $\hat V_3^{3bdy}$, respectively\cite{haldane1983fractional,simon2007generalized}. Thus, the $\hat V_3^{3bdy}$ variational energy of an eigenstate of $\hat V_1^{2bdy}$ (in particular, how close it is to zero) is a good measure for whether said eigenstate resides in $\hat H_{\text{MR}}$. As seen in \Cref{fig:3-body energies}, all of the lowest 1000 (gapped) eigenstates of the Hamiltonian $\hat V_1^{2bdy}$ in the Hilbert space containing the Laughlin ground state (\Cref{fig:3-body energies}a), one-quasihole states (\Cref{fig:3-body energies}b), and two-quasihole states (\Cref{fig:3-body energies}c) have very small $\hat V_3^{3bdy}$ variational energies, with a majority of them having energy lower than 0.001, despite $\hat V_3^{3bdy}$ itself not present in the Hamiltonian. The low-lying eigenstates also have very high overlap with the MR CHS \cite{seesup}.

To see the physical significance of this description, it is instructive to consider the microscopic descriptions of the three types of anyons, labeled $1$, $\psi$, and $\sigma$ following the Ising fusion rule \cite{kitaev2006anyons}:
\begin{align}
\sigma\times\sigma&=1+\psi\label{Ising1}\\
\sigma\times\psi&=\sigma\label{Ising2}\\
\psi\times\psi&=1\label{Ising3}
\end{align}
Microscopically, it can be understood that a $1$-anyon is a localized magnetic flux, a $\psi$-anyon is a magnetic flux bound to a Majorana fermion (MF), and a $\sigma$-anyon is a half-flux \cite{moore1991nonabelions}. Thus, \Cref{Ising1} describes the fusion outcome of combining two half fluxes, while \Cref{Ising3} describes creation/annihilation of a MF pair. These anyons are created from flux insertion in to the ``vacuum", which is the MR ground state of $\hat V_3^{3bdy}$, or the highest density state of $\mathcal H_{\text{MR}}$.

\begin{figure}
\begin{center}
\includegraphics[width=0.85\linewidth]{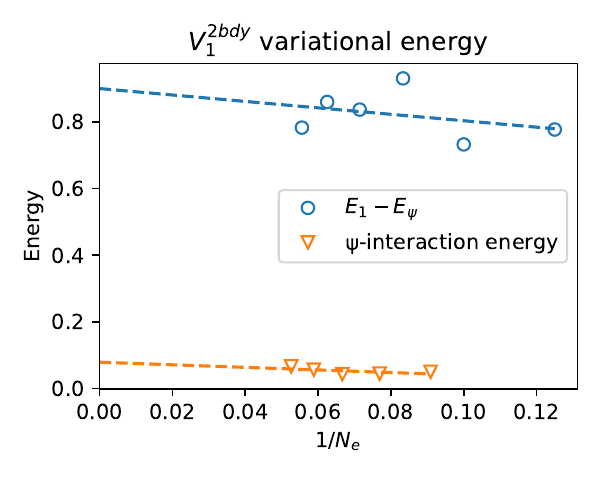}
\caption{Finite-size scaling of difference in self-energy between $1$-quasihole and $\psi$-quasihole (blue) and interaction energy of $\psi$-quasiholes (orange) at filling factor $\nu=1/3$, both calculated for a background $\hat V_1^{2bdy}$ interaction \cite{seesup}. Dotted line shows linear fit. 
}
\label{fig:MR V1 energy}
\end{center}
\end{figure}

All low-lying excitations in the Laughlin phase with only short-range two-body interactions are quantum fluids of these three types of anyons (also see \cite{seesup}). To understand the nature of the Laughlin ground state itself, it is then helpful to consider the dynamics of these three anyons under $\hat V_1^{2bdy}$ interaction. The data from Ref. \cite{xu2025dynamics} has shown that between $\sigma$-quasiholes and $\psi$-quasiholes, $\hat V_1^{2bdy}$ energetically favours a $\psi$-quasihole over its splitting into two $\sigma$-quasiholes. All that is left is to compare $\hat V_1^{2bdy}$ energies of $\psi$-quasiholes and $1$-quasiholes, which we show in \Cref{fig:MR V1 energy}a. Here, working in the MR CHS with two additional fluxes, we place two quasiholes on the opposite poles of the sphere and calculate its $\hat V_1^{2bdy}$ variational energy. Since the size of the sphere is proportional to the number of Landau orbitals $N_o$,  varying the system size is equivalent to varying the separation distance between the quasiholes at the two poles. In general, the variational energy of a two-anyon state is the sum of the anyons' creation energies and their interaction energy. The latter is expected to vanish as their spatial separation goes to infinity \cite{xu2025dynamics}.
We thus examine the extrapolation of these variational energy in the thermodynamic limit, which shows that $\psi$-quasiholes are strongly favoured (having lower variational energy). Thus, of the three quasihole species, $\psi$-quasiholes are the most energetically favoured by $\hat V_1^{2bdy}$. 

That $\hat V_1^{2bdy}$ energetically favours the creation of $\psi$-quasiholes can also be understood from a microscopic argument as follows. In the MR wavefunction, the Pfaffian term can be viewed as a pairing of electrons \cite{moore1991nonabelions,Read1996}. When two or more fluxes are added to the MR ground state, two $1$-anyons can be transmuted into two $\psi$-anyons following the fusion rule $\psi\times\psi=1$. During the process, a MF is bound to each of the $1$-type quasihole. The reverse process, transmuting $\psi$ to $1$, corresponds to the unbinding of a two MFs from two magnetic fluxes. Binding a MF to a magnetic flux requires the splitting of a pair of bound electrons -- analogous to the breaking of Cooper pairs in topological superconductors \cite{cooper1956bound,bardeen1957microscopic}. Since $\hat V_1^{2bdy}$ energetically punishes electron pairing, at lower filling factors the total energy is lowered by breaking up electron pairs, which forms $\psi$-anyons. 

Given that the Laughlin ground state is the zero-energy ground state of $\hat V_1^{2bdy}$ from the insertion of $N_e$ fluxes into the MR ground state (e.g. the vacuum), it must follow that the Laughlin state is a fluid of $N_e$ $\psi$-quasiholes. This conclusion also holds even if we take into account the anyon-anyon interactions. Indeed, at the Laughlin filling factor $\nu=1/3$, the density of $\psi$ is relatively high and effective pair-wise interaction between anyon, which depends on both their spatial separation and the underlying electronic interaction \cite{xu2025dynamics}, potentially affects the dynamics of the system. However, we found that in this case, at the filling factor $\nu=1/3$, $\psi$-quasiholes are only weakly interacting: their interaction energy per pair is two orders of magnitude lower than the self-energy difference between $1$- and $\psi$- quasiholes (see \Cref{fig:MR V1 energy}) \cite{seesup}.

It is important to note that the low-lying gapped excitations in \Cref{fig:3-body energies} also contain a mixture of all three types of MR quasiholes, in particular the non-Abelian $\sigma$-quasiholes. Thus, while the ground state is an Abelian manifold gapped by $\hat V_1^{2bdy}$, non-Abelian excitations are potentially accessible by creating Laughlin neutral excitations (a quasihole-quasiparticle pair \cite{yang2013analytic}), effectively bringing the state outside the nullspace of $\hat V_1^{2bdy}$. If only a small number of neutral excitations are created, such that the $\hat V_1^{2bdy}$ energy is low enough, we have seen in \Cref{fig:3-body energies} that the resulting state still mostly lies within $\mathcal H_{\text{MR}}$. This means that low-lying Laughlin neutral excitation states can still be described in terms of MR quasiholes (see \Cref{fig:Illustration}). Interestingly, dressing the Laughlin quasiholes with neutral excitations can result in fractionalized Laughlin quasiholes, each with charge $e/6$ \cite{trung2021fractionalization}. We will see below that these fractionalized Laughlin quasiholes can obey \emph{non-Abelian}  statistics.

\textit{Statistics of fractionalized Laughlin quasiholes} -- Having established the low-lying (including gapped) states of the Laughlin phase are quantum fluids of non-Abelian anyons, we now consider the Laughlin quasiholes. The Laughlin model wavefunction with $N_e$ electrons and $N_q$ quasiholes can be written, up to a normalization constant, as \cite{laughlin1983anomalous}
\begin{equation}
\label{Laughlin quasiholes}
\prod_{j=1}^{N_q}\prod_{i=1}^{N_e}(z_i-a_j)\prod_{i<j}(z_i-z_j)^3e^{-\sum_{i}|z_i|^2/4\ell_B^2}
\end{equation}
where $z_i=x_i+iy_i$ is the holomorphic variable representing the spatial co-ordinates of the $i$-th electron and similarly $a_j$ is the holomorphic variable representing the spatial co-ordinates of the $j$-th quasihole \cite{seesup}. Each factor $\prod_i(z_i-a_j)$ represents a flux insertion at $z=a_j$. In the MR CHS, this factor corresponds to addition of a $1$-anyon quasihole at the same position due to parity conservation. Thus, the Laughlin quasihole is a $1$-anyon when viewed in the MR CHS picture. When $N_q\geq 2$, the process of transmuting two $1$-anyons to two $\psi$-anyons is possible and energetically favoured by $\hat V_1^{2bdy}$; however no two Laughlin quasiholes can become a pair of $\psi$-anyon because the maximum number of $\psi$-anyons in a finite system is $N_e$. This can be understood as a finite-size cut-off to the countings of the MF modes which can be seen, for example, from the entanglement spectrum \cite{li2008entanglement,sohal2020entanglement,seesup} or first-quantized wavefunction~\cite{Read1996}. Thus, while the Laughlin ground state is a fluid of $\psi$-anyons, all Laughlin quasiholes are $1$-anyons.

A $1$-anyon quasihole in the MR state can be fractionalized into two $\sigma$-anyons. Thus, this picture implies the possibility of fractionalizing the Laughlin quasiholes into non-Abelian quasiholes. Before discussing the physical realization of such a process (in the next section), we provide here an argument for their non-Abelian statistics. The state with $N_q$ Laughlin quasiholes occurs when $N_e+N_q$ magnetic fluxes are added to the MR ground state, which creates $2(N_e+N_q)$ MR quasiholes. Without short-range electron-electron interaction, specifying each of these MR quasiholes results in a $2^{N_e+N_q-1}$-fold degeneracy. However in the presence of $\hat V_1^{2bdy}$ or Coulomb interaction, if we only have $2N_q$ local potentials to trap $2N_q$ of these quasiholes, the short-range interaction will drive the formation of $N_e$ $\psi$-anyons by fusing $N_e$ pairs of quasiholes. This process results in the breaking of a $2^{N_e}$-fold degeneracy into the unique Laughlin ground state. Thus if the remaining localized $2N_q$ quasiholes are well separated and non-interacting, they give a degeneracy of $2^{N_q-1}$ , which is the expected degeneracy of $2N_q$ non-Abelions in the Ising anyon model.

In a realistic system, the nature of the ground state of the combination of $\hat V_1^{2bdy}$ and potential pins discussed above is a little more complex as there exist other possibilities. Instead of each potential pin trapping a $\sigma$-quasihole, another possible scenario is that the ground state consists of a superposition of different occupations of $N_q$ $1$-quasiholes in $2N_q$ slots. In this case, each $1$-quasihole is a Laughlin quasihole and the entire system remains in the Laughlin phase. Thus, the quasihole manifold is Abelian, albeit with larger degeneracy (this is analogous to the quasiholes in two-component Halperin state \cite{halperin1983theory,lopez2001effective}). In view of the competition between these different outcomes, it is important to note that while it is theoretically possible to fractionalize $N_q$ Laughlin quasiholes into $2N_q$ non-Abelions, it must be done with carefully chosen potential pin profile as there also exist other possibilities whose ground-state manifolds are degenerate but Abelian.

\textit{Energetics of local potentials} -- In practice, one can attempt to fractionalize  $N_q$ Laughlin quasiholes by applying an electrostatic potential consisting of $2N_q$ local ``anti-dot''. Each anti-dot effectively traps a quasihole by repelling electrons around it. Since the three species of MR quasiholes have different local density profile \cite{trung2025long,seesup}, they respond differently to a given potential profile. Recall that the goal is to bring down the energy of the gapped excitations containing the non-Abelian $\sigma$-quasihole (effectively dressing each Laughlin quasihole with neutral excitations), the key is then to pick a potential profile that strongly favours $\sigma$-quasiholes over $1$-quasiholes and $\psi$-quasihole.  A phase transition into non-Abelian quasiholes could then be induced if the one-body trapping potential is strong enough to drive the energies of the $\psi$- and $1$-quasiholes higher than that of the $\sigma$-quasiholes: the trapping potentials can be strong enough to attract the $\sigma$-quasiholes to overcome the strong attraction binding them in the $1$-quasiholes. 

The effect of different electrostatic potential profiles on trapping the MR quasiholes within $\mathcal H_{\text{MR}}$ has been discussed in Ref. \cite{storni2011localized}. Here, we investigate the effects of trapping potential on fractionalizing Laughlin quasiholes.
To that end, we perform exact diagonalization (ED) of a general Hamiltonian of the form:
\begin{equation}
\label{12body}
\hat H=\hat V_1^{2bdy} +\lambda \hat V_{N}^{1bdy}
\end{equation}
where $\hat V_1^{2bdy}$ is the model Hamiltonian of the Laughlin state and $\hat V_{N}^{1bdy}=\sum_{i=1}^N\hat V_{\mathbf r_i}$ is the one-body electrostatic potential consisting of $N$ localized anti-dots centered at $\mathbf r_i$ (for $i=1,...,N$), each with a profile of a Gaussian function of the form:
\begin{equation}
\label{gaussian pin}
V_{\mathbf r_i}(\mathbf r)=\frac{1}{\sqrt{2\pi a^2}}e^{-\frac{|\mathbf r-\mathbf r_i|^2}{2a^2}}
\end{equation}
Here we are using the spherical geometry, where the number of orbitals at the lowest Landau level (LLL) is $N_o=2S+1$ where $2S$ is the strength of the monopole at the center of the sphere \cite{haldane1983fractional,greiter2011landau}. In \Cref{gaussian pin}, $a$ parametrizes the Gaussian profile's width. We numerically study the effect of the interplay between the two terms in \Cref{12body}, particularly in the large $\lambda$ limit which may induce a strong mixing between the Laughlin state and the gapped excitations. The choice of the Gaussian width is important: we want the pin to be localized in real space (i.e. $a$ cannot be too large), yet if the pin is too narrow (becoming a Dirac delta in the limit $a\to0$) it will potentially not induce a phase transition because the $1$-quasihole has a lower electron density near the center \cite{seesup}. In the main text we present the analysis for $a=1$ and numerical results for a few other values of $a$ are presented in the supplementary.

\begin{figure}
\begin{center}
\includegraphics[width=0.9\linewidth]{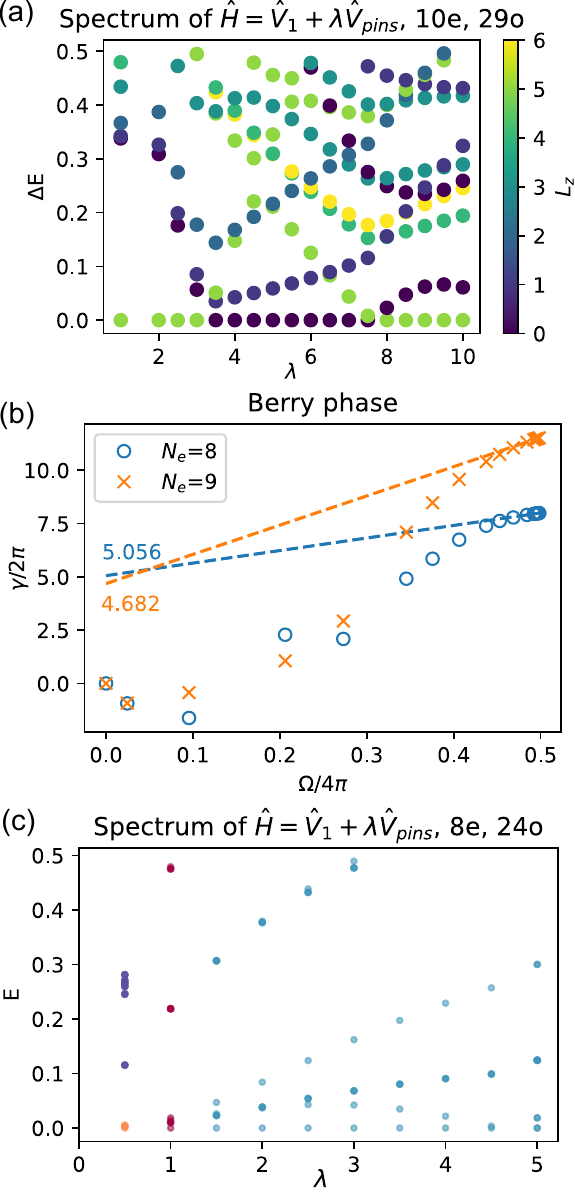}
\caption{(a) Spectrum of \Cref{12body} at increasing $\lambda$ when two potential pins are applied to a system with one quasihole. The color of each point denotes the $L_z$ sector of the corresponding eigenstate. (b) Berry phase obtained by exchanging two potential pins along a circular path as a function of the enclosed solid angle. The colored text inset shows the statistical phase extracted from this data \cite{seesup}. (c) Spectrum of \Cref{12body} at increasing $\lambda$ when four potential pins, placed at the vertices of a regular tetrahedron, are applied to a system with two quasiholes. Data point colors are qualitative indicators explained in the main text.}
\label{fig:Laughlin pin spectrum}
\end{center}
\end{figure}

We first consider the case $N_q=1$ and apply $N=2$ potential pins. To maximize the distance between these two pins, we place them one at the north pole and the other at the south pole of the sphere. This also has the benefits of conserving angular momentum $L_z$, thus using the $L_z$-eigenstate basis the matrix of \Cref{12body} is reduced to a block-diagonal form. At $\lambda=0$, the ground state degeneracy across all $L_z$ sectors is $N_e(N_e+1)/2$, which reflects the bosonic counting of the quasihole \cite{yang2021statistical}. As $\lambda$ increases, rotational symmetry on the sphere (translational symmetry on the plane) is broken and many of these degenerate states are lifted. In the resulting ground state, ``quasihole-like" regions of electron density depletion form around the pins. Here, one of the following two possibilities may happen: the ground state may be two-fold degenerate or unique. The former corresponds to a scenario where the Laughlin quasihole remains unfractionalized (the degeneracy coming from the quasihole being in the superposition of two positions), while the latter happens when it is fractionalized. In our numerical setup, in the first scenario the ground state occurs at $L_z=\pm Ne/2$, and in the second scenario it is at $L_z=0$. Indeed, in our numerics we observe a finite interval $\Lambda$ where the ground state is exactly unique for $\lambda\in\Lambda$ (see \Cref{fig:Laughlin pin spectrum}a, where $\Lambda\sim[3.5,7]$). This length of this interval decreases with system sizes but does not seem to vanish in the thermodynamic limit with a finite pin strength~\cite{seesup}.

The fact that a Laughlin quasihole can only be fractionalized with wide potential pins reflects the qualitative and quantitative difference in the internal structure of the quasiholes. Here it becomes important that each quasihole is not a point particle, but a wavepacket with a finite size \cite{wu2014braiding,johri2014quasiholes,li2022anyonic} and an internal geometric degree of freedom \cite{Umucalllar2018,Macaluso2019,Comparin2021,trung2023spin,ji2024universal}.
The local charge density around a quasihole consists of a central region of charge deficiency that roughly equal the charge of the quasihole and an oscillating ``tail" \cite{trung2023spin,ji2024universal}. The microscopic detail of this oscillation, which arises from the dipole moment coming from the topological shift, is a characteristic feature that differs for different types of quasiholes. Since different charge densities respond differently to different electrostatic potentials, tuning the profile of the pinning potentials (for example, by using scanning tunneling microscopy (STM) tips \cite{papic2018imaging} with different shapes) can be a powerful tool in manipulating and, in particular, fractionalizing quasiholes within a FQH plateau.

It should also be noted that because we have chosen a localized potential pin profile, a unique ($L_z=0$) ground state here necessarily implies that each potential pin traps a quasihole of charge $e/6$, i.e. one half of a Laughlin quasihole. This is because far away from both pins (i.e. near the equator), the electron density necessarily relaxes to filling factor $\nu=1/3$. There are thus two identical objects each residing at a pole, whose total charge is $e/3$. Each fractionalized Laughlin quasihole may be dressed with neutral excitations, but they must have a finite radius and a total charge of $e/6$ in order to obey locality. These fractionalized quasiholes resemble those studied in Ref.~\cite{trung2021fractionalization}, which are confined under $\hat V_1^{2bdy}$ interaction as their attractive interaction are mediated by ``graviton"-like neutral excitations. However, the strong interaction can be screened due to the scattering of the neutral excitations in \cite{trung2021fractionalization}, and it is energetically favourable for the bulk far away from the pins to relax back to the Laughlin ground state. An analogy can also be drawn here to the string breaking observed in lattice gauge theory, where asymptotic freedom is screened by spontaneous pair creations \cite{schilling2000review,bali2005observation}. This phenomenon results in an upper bound in separation energy, which in our setup can be overcome by strong potential pins favouring fractionalized quasiholes. 

The unique ground state and the emergence of two $e/6$ charges strongly suggest physics beyond the simple composite fermion (CF) picture used to describe the physics at $\nu=1/3$, that the more sophisticated parton picture may be needed~\cite{jain1989incompressible}. In addition to the detailed numerical analysis, we also compared the unique ground state with model wavefunctions of fractionalized Laughlin quasiholes, as well as the CF exciton state, in the supplementary materials \cite{seesup}.

\textit{Numerical evidence of non-Abelian braiding} -- Having established the appropriate potential profiles that are capable of fractionalizing the Laughlin quasiholes, we now demonstrate the potential of using such fractionalized quasiholes for non-Abelian braiding operation. We first look for signature of non-Abelian anyon in the $N_q=1$ case. In the Ising model, despite the fusion space of two $\sigma$-anyons being one-dimensional, their non-Abelian property still manifests as the even-odd effect. In the Moore-Read state, this can be seen as the Abelian braiding phase taking different values in the even (even $N_e$) and odd (odd $N_e$) sectors \cite{Read1992,Bonderson2011,Tserkovnyak2003,Macaluso2019,trung2023spin,trung2025long}. Numerical evidence of this even-odd effect in the case of fractionalized Laughlin quasihole can be seen in \Cref{fig:Laughlin pin spectrum}b. Here, following the methodology of Ref. \cite{trung2023spin}, we compute the Berry phase obtained by exchanging two pins twice along a circle on the sphere as a function of enclosed solid angle $\Omega$. The Abelian braiding phase can be read off from this plot as the $y$-intercept of the linear fitting applied on the data points near $\Omega=2\pi$ \cite{trung2023spin}. The even-odd effect is evident in the jump of $-\pi$ in the braiding phase between the even and odd sectors, in agreement with the behavior of the Moore-Read quasiholes known in literature \cite{Read1992,Bonderson2011,Tserkovnyak2003,trung2023spin}.


We also investigate the signature of non-Abelian braiding in the ground-state degeneracy when two Laughlin quasiholes are fractionalized by four potential pins ($N_q=2$ and $N=2N_q=4$). The expected degeneracy of four fractionalized Laughlin quasiholes following the Ising $\sigma$-anyon is two-fold, contrary to the six-fold degeneracy in the unfractionalized case (two quasiholes choosing two out of four pins). In our numerical simulation, four potential pins are placed at the four vertices of a regular tetrahedron inscribing the sphere, with one vertex at the north pole. Beside ensuring maximum possible separation between any pair of anyons, this configuration also simplifies the numerics due to its $C_3$ rotational symmetry about the $z$-axis \cite{seesup}. Note that in the thermodynamic limit, we expect qualitatively the same physics as discussed above for the $N_q=1$ case: namely, the Laughlin quasiholes are fractionalized within a certain range $\lambda\in\Lambda$, signified by a gap-closing and a change in ground-state degeneracy. Note that the fractionalization only depends on the coupling between the local electron density around each quasihole and the local potential pin profile. Hence, so long as the quasiholes are far enough from each other, fractionalizing one quasihole is a local process, unaffected by the presence of other quasiholes in the system. We thus expect, in particular, that two Laughlin quasiholes can be fractionalized by four wide potential pins, similar to fractionalizing one quasihole with two potential pins discussed in the previous section.

Indeed, one can see in the numerical results in \Cref{fig:Laughlin pin spectrum}c the fractionalization of two quasiholes, as well as evidence of the non-Abelian statistics of the resulting four anyons. We see that as $\lambda$ is increased, there is a strong mixing within the quasihole manifold (exact zero-energy states at $\lambda=0$ and the neutral excitations (the continuum of excited states at $\lambda=0$). Particularly, at $\lambda\sim1$ (red points) the original six-fold degeneracy (yellow points) is split. This signifies a phase transition outside of $\mathcal H_{\text{L}}$. At around $4\leq\lambda\leq5$, we see that the lowest two eigenvalues are very close together compared to the higher eigenvalues. That these two eigenvalues are quasi-degenerate with energy difference quickly decreasing with system sizes suggests genuine topological degeneracy. The energy gap between these two states also increases with increasing $\lambda$, which can be accounted for by fusion channel-dependent anyon-anyon interaction at finite distance \cite{xu2025dynamics}.

\textit{Comments and Outlook} -- It is paradoxical to have anyons other than Laughlin quasiholes in the Laughlin phase without topological phase transition, let alone the non-Abelian ones. However the universal topological physics of the Laughlin phase is only robust below the incompressibility gap. The gapped excitations are generally non-universal, and what we are really proposing here is the following: there are still some universal aspects within the spectrum of non-universal gapped states that can be experimentally accessed. Such gapped excitations are generally ignored in theories only interested in the universal aspects of the topological systems (e.g. topological field theory or category theory where all energy scales are implicitly sent to either zero or infinity). Using local trapping potentials, we show it is possible to induce a phase transition within a ``local bubble" with each Laughlin quasihole fractionalizing into two Moore-Read $\sigma-$ anyons with well defined fusion rules, thereby determining their non-Abelian topological degeneracy. Numerics seems to suggest that when the $\sigma-$anyons are well separated they can be weakly interacting. If this is not true in the thermodynamic limit, braiding such anyons will lead to non-universal dynamical phases. Such non-universal phases are also present for any anyons (e.g. Laughlin quasiholes or Moore-Read quasiholes at $\nu=1/2$) that are not well separated, or when their shapes are deformed \cite{trung2023spin}, but they should not affect the non-Abelian nature of the braiding sequences.  

It is important to note that while the artificial $\hat V_1^{1bdy}$ is used for the Laughlin phase throughout this paper for simplicity (e.g. model wavefunctions can be used to illustrate the physics), all qualitative behaviors of the Laughlin quasihole fractionalization can be reproduced with (screened) Coulomb interaction (most likely for any short-range two-body interactions that supports the Laughlin phase) \cite{seesup}. It is also very useful from an experimental point of view that near the filling factor $\nu=1/3$, the low-lying excitations are always within $\mathcal H_{\text{MR}}$ to a very good approximation with short-range two-body interaction, even if such interaction does not support a gapped MR phase at $\nu=1/2$. This would make the filling factor $\nu=1/3$ filling an extremly promising platform for realizing non-Abelian physics, given that the nature of the filling factor $\nu=5/2$ observed in experiments are still controversial. In fact, considering the realistic Coulomb interaction in the first Landau level in experiments, the (nearly) degenerate non-Abelian quasihole manifold is better stabilized by potential pins at the $1/3$ filling compared to $5/2$ \cite{seesup}. This could be because at $\nu=5/2$, the Coulomb interaction is a strong perturbation to the model three-body interaction, and it is well known the quasihole states near this filling lie almost entirely outside of the MR CHS, a stark contrast to the states realized with realistic interactions near $\nu=1/3$.

While this paper focused on realizing MR quasiholes within the Laughlin phase, the arguments in the paper can be generalized to many other FQH phases within the CHS hierarchy. Some other notably interesting examples includes the Gaffnian state (whose quasiholes have rich non-Abelian braiding properties) or the Read-Rezayi parafermion state (whose quasiholes realize the Fibonacci anyons, a suitable candidate for universal quantum computation). Both of these examples are, in principle, possible to be realized also within the Laughlin filling factor with different choices of electrostatic potential profile. Thus, the results presented here are the first in a series of explorations into novel, potentially more accessible and robust platforms for probing non-Abelian physics in experiments.


\textit{Acknowledgement} -- We would like to acknowledge helpful discussions with Steve Simon, Jainendra Jain, Ajit C. Balram, Yuzhu Wang, Zeno Baccico, Wenqi Yang. This work at Nanyang University, Singapore, is supported by the Singapore Ministry of Education (MOE) Academic Research Fund Tier 1 Grant (No. RG156/24), Singapore Ministry of Education (MOE) Academic Research Fund Tier 3 Grant (No. MOE-MOET32023-0003) “Quantum Geometric Advantage”, and Singapore Ministry of Education (MOE) Academic Research Fund Tier 2 Grant (No. MOE-T2EP50124-0017).

\bibliography{ref}

%
%
\clearpage

\onecolumngrid

\renewcommand{\thesection}{S\arabic{section}}
\renewcommand{\thesubsection}{\thesection.\arabic{subsection}}
\renewcommand{\thesubsubsection}{\thesubsection.\arabic{subsubsection}}
\renewcommand{\thefigure}{S\arabic{figure}}
\renewcommand{\theequation}{S\arabic{equation}}
\renewcommand{\thepage}{S\arabic{page}}
\setcounter{figure}{0}
\setcounter{page}{1}
\setcounter{equation}{0}

\setcounter{secnumdepth}{3}

\begin{center}
{\large \textbf{Supplementary Material for ``Non-Abelian Statistics in Abelian Fractional Quantum Hall Phases Induced by Electrostatic Potential''}}

\vspace{1cm}

\noindent\mbox{%
    \parbox{0.8\textwidth}{%
        \indent 
In this supplementary we present more in-depth discussion of the important concepts discussed in the main texts along with additional numerical data. Aiming for this paper to be self-contained, we present a brief summary of well-known results regarding quantum Hall systems, including Landau level (LL) wavefunctions on the disk and sphere geometries, the explicit model wavefunctions for the Laughlin and Moore-Read (MR) states, and the expressions for their respective model Hamiltonian (two-body and three-body pseudopotentials). We will also present the details of the numerical methods, along with additional numerical results using Coulomb interaction. A second-quantization construction of a model state is outlined, from which we obtain a single state with very high overlap with the ground state of potential pins obtained from exact diagonalization.
    }%
}
\end{center}

\section{Single-particle Landau Level Physics}
\subsection{The Disk Geometry}
The simplest two-dimensional system is that of an electron moving in an infinite plane, also called the disk geometry. The Hamiltonian describing a single electron in a uniform magnetic field $B$ is
\begin{equation}
\label{single particle}
\mathbf H = \frac{1}{2m_e}\left(\hat{\mathbf p}-e\mathbf{A}\right)^2=\frac{\hbar eB}{m_e}\left(\hat a^\dagger \hat a+\frac{1}{2}\right)
\end{equation}
where $m_e$ is the electron mass, $e$ is the magnitude of the electron charge, $\mathbf A$ is the vector potential, and the Hamiltonian is diagonalized by the ladder operators
\begin{align}
\hat a^\dagger &= \frac{1}{\sqrt{\hbar eB}}\left(\pi_x + i\pi_y\right)\label{a dagger}\\
\hat a &=\frac{1}{\sqrt{\hbar eB}}\left(\pi_x - i\pi_y\right)\label{a}
\end{align}
One can verify that $[\hat a^\dagger,a]=1$ as expected of ladder operators. The operators $\hat a^\dagger$ and $\hat a$ respectively raises and lowers the energy levels, which are called Landau levels (LLs). There exists a lowest Landau level (LLL) on which the wavefunctions satisfy:
\begin{equation}
\label{LLL condition}
\hat a |\psi\rangle = 0
\end{equation}
Each LL contains a large degeneracy, which can be resolved by another set of ladder operators:
\begin{align}
\hat b &=\sqrt{\frac{eB}{\hbar}}\left(\tilde R_x-i\tilde R_y\right)\label{eq:guiding center lower}\\
\hat b^\dagger &=\sqrt{\frac{eB}{\hbar}}\left(\tilde R_x+i\tilde R_y\right)\label{eq:guiding center raise}\\
\tilde R_x &= \hat x+\frac{1}{eB}\hat \pi_y\\
\tilde R_y &= \hat y-\frac{1}{eB}\hat \pi_x
\end{align}
Since each of $\hat a^\dagger$ and $\hat a$ (collectively referred to as the \emph{cyclotron} ladder operators) commutes with each of $\hat b^\dagger$ and $\hat b$ (collectively referred to as the \emph{guiding center} ladder operators, a complete basis for the single-particle Hilbert space is given by
\begin{equation}
\label{LL basis}
|n,m\rangle = \frac{1}{\sqrt{n!m!}}\left(\hat a^\dagger\right)^n\left(\hat b^\dagger\right)^m|0,0\rangle
\end{equation}
where the state $|0\rangle$ satisfiles $\hat a|0,0\rangle=\hat b|0,0\rangle = 0$. Proceeding to working only on the LLL, we will omit the LL index $n=0$ and write $|m\rangle\equiv|0,m\rangle$. It is also useful to write down the first-quantized wavefunctions for these states, which requires picking a gauge for the vector potential $\mathbf A$. Using the symmetric gauge $\mathbf A = B/2(-B,B)^{\text{T}}$, the first-quantized wavefunctions of the LLL basis is
\begin{equation}
\label{LLL wavefunction disk}
\phi_m(z)\equiv\langle r|m\rangle = \sqrt{\frac{2^m\ell_B^{2m}}{(2\pi)m!}}z^me^{-|z|^2/4\ell_B^2}
\end{equation}
where $z=x+iy$ and $\ell_B=\sqrt{\hbar/eB}$ is the magnetic length. The form of \Cref{LLL wavefunction disk} means that \emph{any} LLL wavefunction must be of the form $f(z)e^{-|z|^2/4\ell_B}$ where $f(z)$ is a \emph{holomorphic} function (a function of only $z$ and not $\overline{z}$).

\subsection{The Spherical Geometry}
Instead of the infinite plane like in the discussion above, the electron can also be placed onto the surface of the sphere. A uniform magnetic field perpendicular to the surface of the sphere everywhere can be generated by a magnetic monopole of strength $2S\hbar$ placed at the center. Dirac quantization requires that $2S$ be an integer. The Hamiltonian can be written in terms of the (dynamical) angular momentum operator $\hat{\mathbf \Lambda}=\hat{\mathbf r}\times(\hat{\mathbf p}-e\mathbf A)$ as
\begin{equation}
\label{single particle sphere}
\hat H = \frac{1}{2m_e R^2}\hat{\mathbf \Lambda}^2
\end{equation} 
$\Cref{single particle sphere}$ can be diagonalized by defining a set of angular momentum operator:
\begin{align}
\hat{\mathbf L} &= \Lambda + \frac{S}{R}\hat r\label{L}\\
\hat{L}^2 &= \Lambda^2+S^2\label{L^2}
\end{align}
which follow the well-known Lie algebra of the angular momentum operators. Thus, eigenstates of \Cref{single particle sphere} is the simultaneous eigenstates $|l,m\rangle$ of $\hat{L}^2$ and $\hat{L}_z$:
\begin{align}
\hat L^2|l,m\rangle &= \hbar^2l(l+1)|l,m\rangle\label{L2 eigenstate}\\
\hat L_z|l,m\rangle &= \hbar m|l,m\rangle\label{Lz eigenstate}
\end{align}
Here $l$ takes a minimum value of $S$, and we can write $l=S+n$ where the integer $n$ is now the LL index. The eigenvalues of \Cref{single particle sphere} also forms Landau levels, but unlike the plane, the energies of the LLs on the sphere are not evenly spaced. The LL energy varies with the LL index $n$ ($n\geq 0$) as
\begin{equation}
\label{LL sphere}
E_n=\frac{\hbar eB}{2m_eS}\left[(2n+1)S+n(n+1)\right]
\end{equation} 
To write down the first-quantized wavefunction, we pick the gauge
\begin{equation}
\label{lattitude gauge}
A_\varphi = -\frac{S}{eR}\cot\theta
\end{equation}
where the azimuthal angle $\theta$ and polar angle $\varphi$ parametrizes positions on the sphere. The corresponding LLL ($l=S$) wavefunction is
\begin{align}
\phi_{S,m}(\theta,\varphi)\equiv\langle \mathbf r|S,m\rangle&=\sqrt{\frac{(2S+1)!}{(S+m)!(S-m)!}}u^{S+m}v^{S-m}\label{LLL wavefunction sphere}\\
u&=\cos{\frac\theta2}e^{i\varphi/2}\label{spinor u}\\
v&=\sin{\frac\theta2}e^{-i\varphi/2}\label{spinor v}
\end{align}.
\Cref{LLL wavefunction sphere} is related to \Cref{LLL wavefunction disk} by a stereographic projection $z=v/u$, which transforms \Cref{LLL wavefunction sphere} to
\begin{equation}
\label{LLL wavefunction sphere holomorphic}
\phi_{S,m}\propto z^{S-m}u^{2S}
\end{equation}
Thus any LLL wavefunction on the sphere must be of the form $f(z)u^{2S}$ where $f(z)$ is a holomorphic function. Comparing this result with the form of the holomorphic wavefunction on the disk, we can see that there exists a one-to-one mapping from a disk wavefunction to the sphere wavefunction by replacing the ``form factor'' $e^{-|z|^2/4\ell_B^2}\mapsto u^{2S}$ followed by renormalizing the wavefunction.

\section{Model Wavefunctions for FQH Phases}
\subsection{The Laughlin State}
On both the disk and the sphere, any LLL takes the form of a holomorphic function $f(z)$ multiplied by a form factor, which is geometry-dependent. For a many-body wavefunction, a LLL wavefunction can be generally specified by a multivariate holomorphic function $f(z_1,z_2,...,z_{N_e})$, which must also be anti-symmetric in order to be a valid electronic wavefunctions. In writing down the model wavefunctions below, we will only write the holomorphic part and omit the form factor (the Gaussian $e^{-\sum_i|z_i|^2/4\ell_B^2}$ on the disk or $\prod_i u_i^{2S}$ on the sphere). To signify that the wavefunction is missing a form factor and an overall normalization factor, we will use the symbol ``$\sim$'' in place of the equal sign ``$=$''.

Laughlin provided the first ansatz describing an FQH phase at filling factor $1/q$ (where $q$ is an odd integer) as 
\begin{equation}
\label{Laughlin ground state}
\psi_{L}(z)\sim\prod_{i<j}(z_i-z_j)^q
\end{equation}
(Note that here, as with in all many-body wavefunction that follows, $\psi(z)$ is short-hand for $\psi(z_1,z_2,...,z_{N_e})$.) The Laughlin state described in the main text is for $q=3$, which describes the most prominent FQH plateau seen in experiments at filling factor $\nu=1/3$.

Quasiholes of the Laughlin state are gapless excitations obtained from inserting additional magnetic fluxes to the system. The model wavefunctions for a quasihole state can be obtained by multiplying to the ground state a ``vortex'' factor $\prod_i(z_i-a)$, which introduces a zero at position $\mathbf a = (a_x,a_y)$ (here $a=a_x+ia_y$). A state with $N_q$ quasiholes positioned at $\mathbf a_1, \mathbf a_2,...,\mathbf a_{N_q}$ is thus given by
\begin{equation}
\label{Laughlin quasihole}
\psi_{L}(z;a_1,a_2,...,a_{N_q})\sim\prod_{k=1}^{N_q}\prod_{i=1}^{N_e}(z_i-a_k)\prod_{i<j}(z_i-z_j)^q
\end{equation}

\subsection{The Moore-Read State}
The Moore-Read (MR) state describes the filling factor $1/q$ for even $q$. To ensure the wavefunction is anti-symmetric, the Laughlin factor with even power is multiplied by a Pfaffian:
\begin{equation}
\label{MR ground state}
\psi_{\text{MR}}(z)\sim\text{Pf}\left(\frac{1}{z_i-z_j}\right)\prod_{i<j}(z_i-z_j)^q
\end{equation}
here $\text{Pf}(\frac{1}{z_i-z_j})$ denotes the Pfaffian of a matrix whose $(i,j)$-entry is $\frac{1}{z_i-z_j}$. The Pfaffian is defined for an anti-symmetric $N\times N$ matrix $A$, where $N$ is an even integer, as 
\begin{equation}
\label{Pfaffian}
\text{Pf}(A) = \frac{1}{2^nn!}\sum_{\sigma\in S_{N}}\text{sgn}(\sigma)\prod_{i=1}^{N/2}A_{\sigma(2i-1),\sigma(2i)}
\end{equation}
Similar to the Laughlin state, gapless excitations can be added to the MR state by insertion of a magnetic flux. However a novel feature not present in the Laughlin state is that a flux can fractionalize \emph{without an energy cost}. The model wavefunction for two half-fluxes at positions $\mathbf a_1, \mathbf a_2$ is
\begin{equation}
\label{MR quasihole}
\psi_{\text{MR}}(z;a_1,a_2)\sim\text{Pf}\left(\frac{(z_i-a_1)(z_j-a_2)+(z_i-a_2)(z_j-a_1)}{z_i-z_j}\right)\prod_{i<j}(z_i-z_j)^q
\end{equation}
When $a_1=a_2=a$, the Pfaffian in \Cref{MR quasihole} simplifies to
\begin{equation}
\label{MR quasihole Pfaffian fuse}
\text{Pf}\left(\frac{2(z_i-a)(z_j-a)}{z_i-z_j}\right)=\prod_{i=1}^{N_e}(z_i-a)\text{Pf}\left(\frac{1}{z_i-z_j}\right)
\end{equation}

\subsection{Model Hamiltonians}
The Laughlin state is the exact zero-energy state of the model Hamiltonian $\hat V_1^{2bdy}$, which is the projection into states with two electrons having relative angular momentum greater than 1. It has a first quantization form:
\begin{equation}
\label{V1}
V_1^{2bdy}(r_1, r_2)\propto \nabla_1^2\delta^2(r_1-r_2)
\end{equation}

On the other hand, the model Hamiltonian for the Moore-Read state is a three-body interaction of the form
\begin{equation}
\label{V3bdy}
V_3^{3bdy}\propto \nabla_1^4\nabla_2^2\delta^2(r_1-r_2)\delta^2(r_2-r_3)
\end{equation}

\section{Self-energy and Interaction Energy of Moore-Read Anyons}
To investigate the self-energy of each type of MR quasihole and the interaction energy between them, we compute the $\hat V^{2bdy}$ variational energies of different model states within the MR CHS and extract the relevant information. This is the same methodology employed in Ref. \cite{xu2025dynamics} to study interaction between $\sigma$ anyons. In the second quantization, model MR states can be obtained by the Jack polynomial formalism \cite{bernevig2008model,bernevig2008generalized,DiagHam}.

The total quasihole energy of a given state is taken as the difference between variational energies of that state and the MR ground state:
\begin{equation}
\label{E total}
E_{\text{total}} = \langle V^{2bdy}\rangle - \langle V^{2bdy}\rangle_{\text{ground state}}
\end{equation}
where ``ground state'' here refers to the unique highest-density state obtained by Jack polynomial. On the sphere the number of electrons $N_e$ and number of orbitals $N_o$ at which this ground state occurs satisfy $N_o=2N_e-2$, for even $N_e$. On the other hand, quasihole states occurs at $N_o=2N_e-2+n$ where $n>1$ is the number of added fluxes to the ground state and $N_e$ may now be either even or odd. In the odd sector, we take $\langle V_1^{2bdy}\rangle_{\text{ground state}}$ in \Cref{E total} to be the linear interpolation of ground state energies of the two adjacent $N_e$ numbers (see \Cref{fig:LLL Coulomb energy}a). 

Assuming that the given state contains only one anyon species, the total energy in general can be written as the sum of the so-called self-energy (creation energy) and the interaction energy:
\begin{equation}
\label{E components}
E_{\text{total}} = E_{\text{self}} + E_{\text{interaction}}
\end{equation}
where the second term is present only for states with two or more quasiholes. A $1$-quasihole and $\psi$-quasihole can exist in isolation -- by adding one flux to the ground state in the even and odd sector respectively. However, for a farer comparison between their self-energies, we can also consider the even sector with two additional fluxes. The model states can be constructed with one quasihole at the north pole and one quasihole at the south pole using the Jack polynomial formalism with the following root configuration:
\begin{align}
011001100...1100110\label{two 1s}\\
1001100...110011001\label{two psis}
\end{align}
where the ellipsis denote repeating patterns of ``$1100$''.

Using \Cref{E total} and \Cref{E components}, we can also approximate the interaction energy of $\psi$-quasiholes at filling factor $\nu=1/3$ as shown in \Cref{fig:MR V1 energy} in the main text. Here, we take $\langle V^{2bdy}\rangle$ in \Cref{E total} to be the variational energy of the Laughlin state, which is exactly zero if $\hat V^{2bdy}=\hat V_1^{2bdy}$. To extract the interaction energy, we need to approximate the total contribution from the self-energy of $\psi$-quasiholes. We take the self-energy of a single quasihole, $E_\psi$, to be the difference in variational energy between the state given by Jack polynomial with root configuration $100110011..110011$ (an odd number of electrons) minus the ground state variational energy of the MR state interpolated at the same number of electrons. The total interaction energy is then simply the difference in total energy and $N_e$ times $E_\psi$. To compute the interaction energy \emph{per pair}, we assume that the total interaction energy scales with the number of $\psi$-quasiholes, which equals $N_e$. All in all, we have:
\begin{equation}
E_{\psi-\psi}=\frac{1}{N_e}\left(\langle\hat V^{2bdy}\rangle_{\text{Laughlin}} - \langle\hat V^{2bdy}\rangle_{\text{Moore-Read}}-N_e\times E_\psi\right)
\end{equation}

\section{Electron density of Moore-Read quasihole states}
\begin{figure}
\begin{center}
\includegraphics[width=0.5\linewidth]{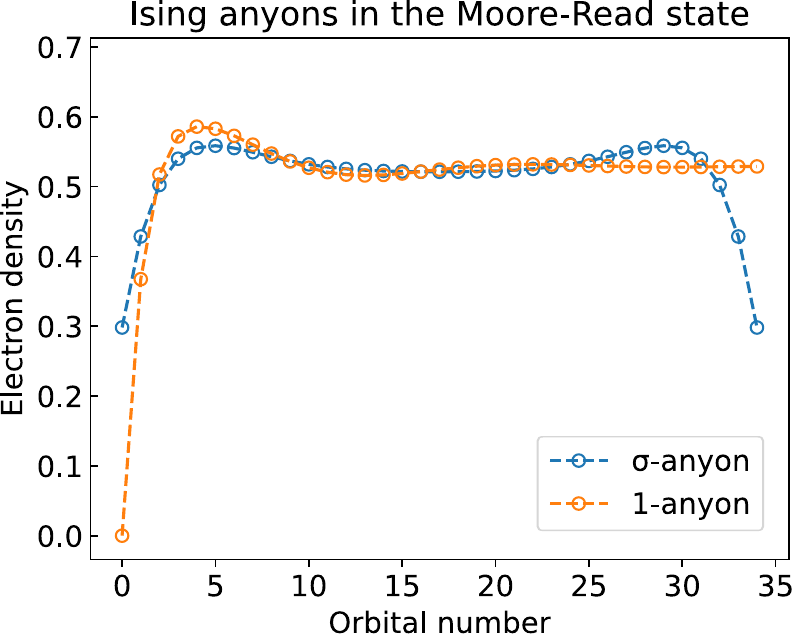}
\caption{Average orbital occupation of the MR state with one $1$-quasihole (orange) and the MR state with two separated $\sigma$-quasihole (blue).}
\label{fig:anyon electron density}
\end{center}
\end{figure}

Since one-body potential couples to the electron density, it can be instructive to look at the electron density profile of the $1$-quasihole and $\sigma$-quasihole state. Both states can be constructed by Jack polynomials with $N_o=2N_e-1$, for even $N_e$, with the following root configuration
\begin{align}
0110011001100... \text{\hspace{30pt}($1$-quasihole)}\label{eq:1 quasihole root}\\
101010....10101 \text{\hspace{30pt}($\sigma$-quasihole)}\label{eq:sigma quasihole root}
\end{align}
where the ellipsis in \Cref{eq:1 quasihole root} denotes repeating ``1100", whereas in \Cref{eq:sigma quasihole root} it denotes repeating ``10''. \Cref{eq:1 quasihole root} gives a $1$-quasihole residing at the north pole, and \Cref{eq:sigma quasihole root} gives two $\sigma$-quasihole, one at the north pole and the other at the south pole. Since both states are eigenstates of $\hat L_z$, their real-space electron density can be inferred from the avarage occupation of each orbital (see \Cref{fig:anyon electron density}).

From the electron density in \Cref{fig:anyon electron density}, it can be seen that if a potential pin punishes only one orbital, which achieved by a Gaussian profile with very small width ($a\ll \ell_B$), then it would trap a $1$-anyon rather than a $\sigma$, since the density at the first (left-most) orbital of the $1$-quasihole state is much lower than that of the $\sigma$-quasihole. Since we aim for the $\sigma$-quasihole to have lower-energy, it is expected that this can only be achieved with wider potential pins, which punishes multiple orbitals around its center. In principle, based on the difference in electron density of the two states in \Cref{fig:anyon electron density} one may be able to design an ``ideal" pin potential profile to strongly favour the $\sigma$-quasiholes, although it is possible that such an ideal profile is difficult to realize in experiment. The finite-width Gaussian profile presented in this paper shows an example that is both realistic and possible of producing $\sigma$-quasihole ground state.

\section{Details on Numerical Calculations}
\subsection{General Hamiltonian}
All numerical results in this paper are obtained from diagonalizing, on the spherical geometry, the Hamiltonian of the general form:
\begin{equation}
\label{12body gen}
\hat H = \hat V^{2bdy} + \lambda\hat V^{1bdy}
\end{equation}
where $\hat V^{2bdy}$ is a general two-body interaction, which can be decomposed as the sums of two-body Haldane pseudopotentials. The one-body potential $\hat V^{1bdy}$ has the form
\begin{equation}
\label{1body gen}
\hat V^{1bdy} = \int d\Omega V(\theta,\varphi)\hat\rho(\theta,\varphi)
\end{equation}
where $\rho(\theta,\varphi)$ is the density operator. $\lambda$ in \Cref{12body gen} denotes an overall strength of the pinning potential compared to the two-body term. We work in the computational unit with $\hbar=c=e=B=1$ and take the LLL limit ($m_e\to\infty$). In this limit the wavefunctions in \Cref{LLL wavefunction sphere} forms a basis for the Hilbert space.

Using the basis for the LLL on the sphere in \Cref{LLL wavefunction sphere}, the projection of $\hat V^{1bdy}$ onto the LLL can be written in second-quantized form as
\begin{equation}
\label{1body second quantized}
\hat P_{LLL}\hat V^{1bdy}\hat P_{LLL}=\sum_{m_1,m_2}v_{m_1,m_2}|S,m_1\rangle\langle S,m_2|
\end{equation}
where $\hat P_{LLL}$ denotes the LLL projection operator and each of $m_1$ and $m_2$ goes from $-S$ to $S$. If the potential profile $V(\theta,\varphi)$ is rotationally symmetric, i.e. $V(\theta,\varphi)=V(\theta)$, then the matrix is diagonal: $v_{m_1,m_2}=v_m$ for $m_1=m_2=m$. To model a potential pin at the north pole, we can choose $v_m$ such that
\begin{equation}
\label{orbital pin coefficients}
v_m=\begin{cases}
1&m>S-k,\\
0&m\leq S-k
\end{cases} 
\end{equation}
In this supplementary, we will refer to this type of potential as the ``orbital'' pin as it equally punishes the first $k$ orbitals around the north pole. 

The potential centered at the north pole in \Cref{1body second quantized} can be transformed such that it is instead centered at some point $(\theta_0,\varphi_0)$ by the operator $\hat R(\theta_0,\varphi_0)=e^{i\varphi_0\hat L_z}e^{i\theta_0 \hat L_y}$ (rotation by $\theta_0$ around the $y$-axis followed by the rotation by $\varphi_0$ around the $z$-axis). The resulting matrix element is then given by
\begin{equation}
\label{rotated 1bdy}
\sum_{m_1,m_2,m}v_m|S,m_1\rangle\langle S,m_1|\hat R(\theta_0,\varphi_0)|S,m\rangle\langle S,m|\hat R^\dagger(\theta_0,\varphi_0) |S,m_2\rangle\langle S,m_2|
\end{equation}
The terms of the form $\langle S,m|\hat R|S,m'\rangle$ is the Wigner-D matrix whose analytical expression is known.

\subsection{Tetrahedral Symmetry}
Having more than two potential pins breaks rotational symmetry of the system and mixes the different $L_z$ sectors. Thus, the system size required in ED is potentially enlarged. However, with the choice of placing four potential pins at the vertices of a regular tetrahedron, it turns out that the overall matrix of \Cref{12body gen} is still block-diagonal in the $L_z$ basis. This is particular due to $C_3$ symmetry from rotation by $2\pi/3$ about the $z$-axis. This symmetry is characterized by a quantum number $T$ which corresponds to the eigenvalue of the rotation operator $\hat R_{2\pi/3} = e^{i\frac{2\pi}{3}\hat L_z}$ operator:
\begin{equation}
\label{C3 quantum number}
e^{i\frac{2\pi}{3}\hat L_z}|T\rangle = e^{\frac{i2\pi T}{3}}|T\rangle\text{, for $T=0,1,2$}
\end{equation}
Since $\hat R_{2\pi/3}$ commutes with both the Hamiltonian \Cref{12body gen} and $\hat L_z$, the Hamiltonian matrix is block diagonal in the $L_z$ basis, consisting of three blocks each corresponding to one of the three values for $T$. Thus, by separating the Hilbert space basis into three groups each with $\langle L_z\rangle\equiv T$ (mod 3), the Hamiltonian can be diagonalized within each basis separately. This greatly reduces the computational cost for ED.

\subsection{Gaussian Potential Profile}
Still using model Hamiltonian $\hat V_1^{2bdy}$, we consider the effect of using potential pin with a profile of a Gaussian function of the form:
\begin{equation}
\label{gaussian}
V(r)=\frac{1}{\sqrt{2\pi a^2}}e^{-\frac{r^2}{2a^2}}
\end{equation}
To map this onto the sphere, we compute the coefficient in the LLL basis $v_m = \langle m|\hat V|m\rangle$ (where $|m\rangle\equiv|0,m\rangle$ is the LL orbital on the disk given in \Cref{LL basis}), and use the same coefficients to construct the one-body operator using the LLL basis on the sphere as
\begin{equation}
\label{1bdy on sphere}
\hat V^{1bdy} = \sum_{m=0}^{2S+1}v_m|S,S-m\rangle\langle S,S-m|
\end{equation}
which gives a Gaussian-profile pin at the north pole. To make the pin centered at any other point on the sphere we apply the rotation operators in the same manner as \Cref{rotated 1bdy}.

\begin{figure}
\begin{center}
\includegraphics[width=\linewidth]{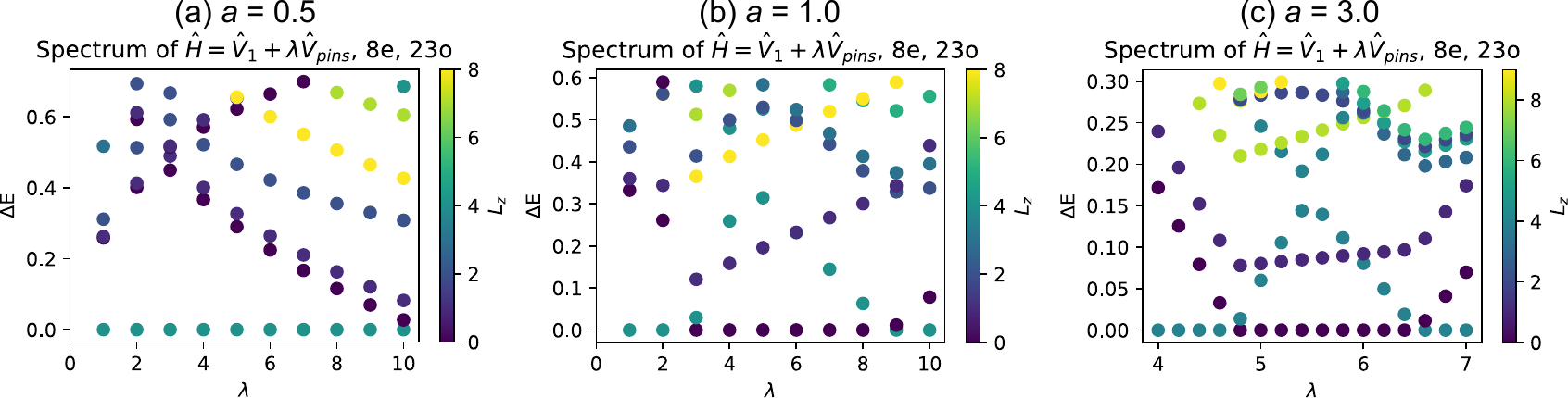}
\caption{Spectrum of \Cref{12body gen} where $\hat V^{2bdy} = \hat V_1^{2bdy}$ and $\hat V^{1bdy}$ consists of two Gaussian pins located at the north and south poles. Each plot shows a different pin size: (a) $a=0.5$ (b) $a=1.0$, (c) $a=3.0$.}
\label{fig:Gaussian different a}
\end{center}
\end{figure}

\section{Ground state properties and model states}
\subsection{Ground state described by Moore-Read quasiholes}
Working with a set up consisting of one quasihole and two potential pins, we are interested in the cases where the ground state is unique and resides at $L_z=0$ sector. Simple inspection on numerical data shows that the choice of the $a$ parameter (Gaussian width) is important. \Cref{fig:Gaussian different a} shows the spectrum for several values of $a$. When $a$ is small, potentially there is no phase transition into a unique ground state, unless the pin strength is very large. (in the limit $a\to0$ it is known that the Laughlin ground state is robust~\cite{rezayi1985incompressible}.) On the other hand, for larger $a$ the range of the interval $\Lambda$ with unique ground state decreases.

\begin{figure}
\begin{center}
\includegraphics[width=0.45\linewidth]{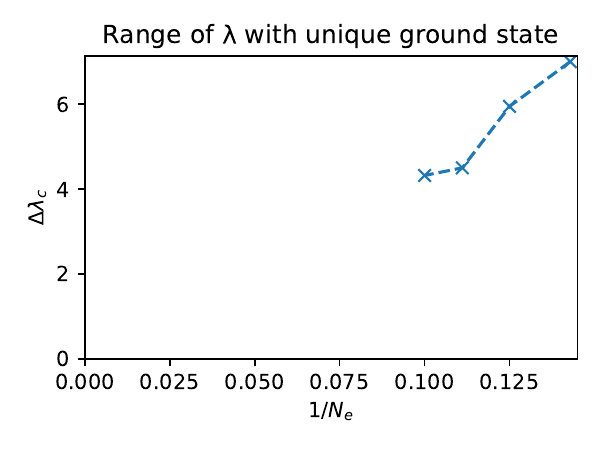}
\caption{Finite size scaling of the length of the interval $\Lambda$ within which the ground state with one quasihole and two pins is unique. Here each pin takes a Gaussian profile with $a=1.0$.}
\label{fig:lambda scaling}
\end{center}
\end{figure}

We take $a=1.0$ (\Cref{fig:Gaussian different a}b) as a sample case to study the nature of the unique ground state. A unique ground state is achieved within a certain interval $\Lambda$. The finite-size scaling of the length of this interval is shown in \Cref{fig:lambda scaling}, which shows it is potentially finite in the thermodynamic limit. The most important thing to note is that the ground state lies almost entirely outside the Laughlin CHS, but almost entirely \emph{within} the Moore-Read CHS, as seen from the wavefunction overlap. We construct a basis for each CHS by orthonormalizing all the corresponding Jack polynomials, and the cumulative overlap within a given Hilbert space spanned by $\{|\phi_0\rangle,|\phi_1\rangle,...,|\phi_d\rangle\}$ is calculated as
\begin{equation}
\label{cumulative overlap}
\left(\sum_{i=1}^{d}|\langle\psi|\phi_i\rangle|^2\right)^{1/2}
\end{equation}
where $|\psi\rangle$ denotes the state being studied.

\subsection{Model state for fractionalized Laughlin quasihole}
An overlap more than 0.99 is a non-trivial result -- indeed, while the Moore-Read CHS is large compared to the Laughlin CHS, it is much smaller than the full many-body Hilbert space of the lowest Landau level (LLL). Furthermore, we can do better by narrowing down the number of basis states to just one. Our construction of this model state is done in second-quantized basis and reveals the physics both near the pins and far away within the bulk of the electron gas.

To construct a model state with $N_e$ electrons, we start by constructing an $(N_e-2)$-electron \emph{Haffnian} state~\cite{green2001strongly} that has a root configuration 0000101000...1010001010000, where the ellipsis denotes repeating pattern of 101000. Note that this root configuration satisfies the $(2,6)$-admissibility rule, as expected of a Haffnian state. We obtain this state by diagonalizing the model three-body Hamiltonian $\hat V_H=\hat V_3^{3bdy}+V_5^{3bdy}+V_6^{3bdy}$ using the monomial squeezed from the above root configuration, from which a \emph{unique} zero-energy ground state is obtained. Then, we add two electrons by applying the operator $\hat c_{S-2}^{\dagger}c_{-S+2}^{\dagger}$ where $\hat c_{m}^\dagger$ is creates an electron at orbital $m$ on the sphere. The state $|\phi_0\rangle$ obtained from this procedure is the unique state that satisfies:
\begin{equation}
\label{eq:Haffnian bulk}
\hat V_H\hat c_{S-2}\hat c_{-S+2}|\phi_0\rangle=0
\end{equation}
Next, we create the remaining basis states $|\phi_1\rangle$, $|\phi_2\rangle$,... by taking the monomials squeezed from $0010101000101000...1010001010100$ that doesn't appear in the monomial expansion of $|\phi_0\rangle$.  The resulting set of states $\{|\phi_0\rangle,|\phi_1\rangle,...\}$ forms an orthonormal basis, and we take our model state to be the ground state of $\hat V_1^{2bdy}$ within this subspace.

Before comparing this model state with the unique ground state of two Gaussian pins, let us discuss the physics implied by this construction. The state $|\phi_0\rangle$ has the property that it is Haffnian in the bulk far away from the two poles. Thus, in the thermodynamic limit, the bulk filling factor is $\nu=1/3$, as expected from locality principle. It has been shown that the Haffnian CHS indeed contains the fractionalized Laughlin quasiholes, whose interaction are mediated by Laughlin neutral excitations~\cite{trung2021fractionalization}. $|\phi_0\rangle$ on its own has very high variational energy with respect of $\hat V_1$ since every neutral excitation is punished (note that in the root configuration, every group of ``101000" denotes a pair of quasielectron-quasihole). This energy is potentially softened by the addition of $|\phi_1\rangle$, $|\phi_2\rangle$... which generally contain fewer neutral excitations.

In the above construction, one electron is added to the third orbital from each pole (angular momenta $S-2$ and $-S+2$, respectively). These two positions are chosen to minimize the electron density at the two poles (center of the potential pins) while making sure that the root configuration still satisfies the \emph{Moore-Read} admissibility rule (the $(2,4)$-admissibility rule). The Haffnian admissibility rule can be violated near the two poles as a result of the potential pins. In the resulting root configuration, the first few digits from the left (0010101000...) represent the physics of the quasihole at the north pole, which can be understood both from the perspective of a Laughlin state (using (1,3)-admissibility rule) or the Moore-Read state (using (2,4)-admissibility rule). From a Laughlin perspective, the object at the north pole is a half-quasihole dressed by a neutral excitation, which has a total electrical charge of $e/6$. From a Moore-Read perspective, this root configuration denotes a $\sigma$-quasihole (represented by 1010101...) and two $1$-quasihole (represented by the two leading 0's). Thus, this object is a fusion of one $\sigma$ with two bosons and has a total \emph{topological charge} $\sigma$, which is non-Abelian in the Ising anyon model.

\begin{figure}
\begin{center}
\includegraphics[width=0.75\linewidth]{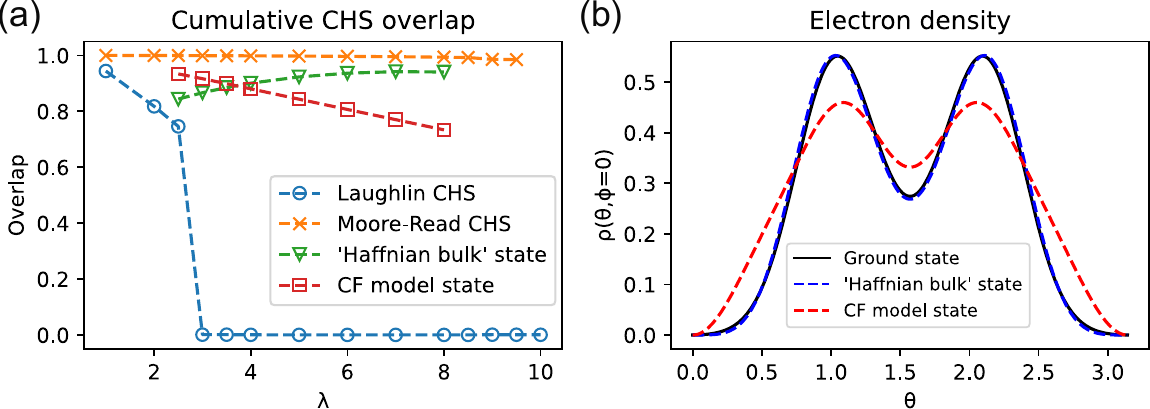}
\caption{(a) Overlap as a function of pin strength of the true ground state of \Cref{12body gen} with two Gaussian pins ($a=1$) with the different Hilbert spaces and model states. (b) The electron density along a latitude of the ground state at $\lambda=7.0$ (black solid line) compared to those of the two model states (dashed lines). Data shown here are computed on a system with 8 electrons and 23 orbitals.}
\label{fig:overlap}
\end{center}
\end{figure}

\subsection{Comparison with composite fermion trial state}
We can see from the overlap shown in \Cref{fig:overlap}a this model state performs very well, with overlap above 0.93 with the true ground state at large pin strength. As a simple benchmark, we compare this model state with a composite fermion (CF) model state consisting of one CF hole one neutral excitation at the lowest $\Lambda$-level. The overlap of the true ground state with this CF model state decreases rapidly with increasing $\lambda$. Our model state also performs extremely well in terms of real-space electron density (see \Cref{fig:overlap}b).

\begin{figure}
\begin{center}
\includegraphics[width=0.75\linewidth]{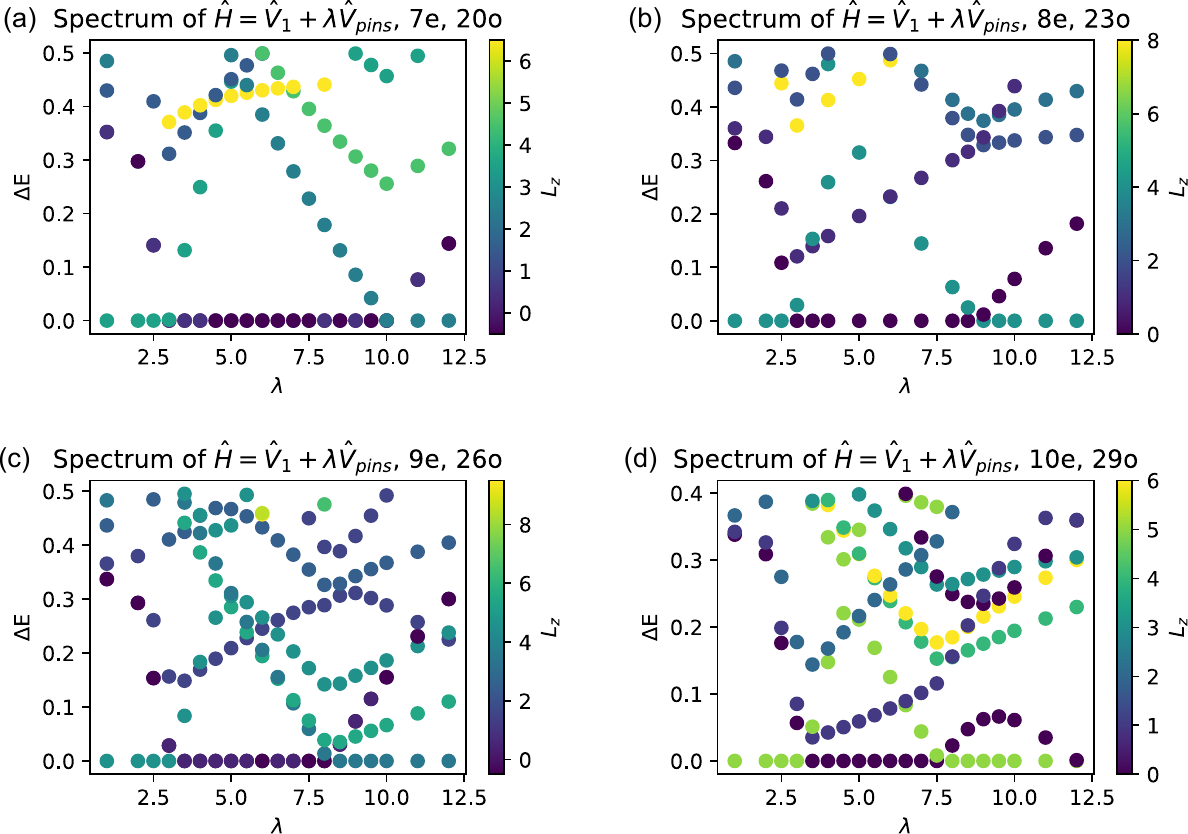}
\caption{Spectrum flow of $\hat V_1^{2bdy}+\hat V_{\text{pins}}$ where $\hat V_{\text{pins}}$ consists of two Gaussian pins ($a=1.0$) placed at the north and south poles, for a system with one Laughlin quasihole with different sizes: (a) 7 electrons (b) 8 electrons (c) 9 electrons (d) 10 electrons.}
\label{fig:different system sizes}
\end{center}
\end{figure}

It should also be noted that the unique ground state can only be observed with an even number of electrons. When the number of electrons, $N_e$, is odd, all eigenstates of the Hamiltonian reside in the half-integer angular momentum sectors. Then, analogous to the discussion above, the ground state of \Cref{12body} resides in the $L_z=\pm N_e/2$ for $\lambda\notin\Lambda$ and $L_z=\pm 1/2$ for $\lambda\in\Lambda$.  (see \Cref{fig:different system sizes}). We can be certain a phase transition has occurred despite no change in the ground state degeneracy. When $\lambda\in\Lambda$, the fractionalized Laughlin quasihole state still exhibits a two-fold degeneracy coming from the asymmetric distribution of an odd number of MFs on an inversion-symmetric system \cite{seesup}. This symmetry-protected degeneracy can be best understood by mapping the spherical geometry onto the cylinder via a conformal stereographic projection, which is the experimentally relevant geometry with two counter-propagating edges. The two degenerate states with fractionalized Laughlin quasiholes are related to the mirror symmetry with the two fractionalized quasiholes located at the two edges. Thus, it is important to note the even-odd effect in finite systems and the symmetry of the system when considering the ground state degeneracy to distinguish this ``symmetry-protected" degeneracy (broken when symmetry is broken) from the non-Abelian degeneracy (unaffected by symmetry).

The comparison here shows that the unique ground state with $e/6$ charge at each pin is unlikely captured by the physics of simple CF states (e.g. product states of composite fermions). A linear combination of many CF product states (i.e. single and multiple CF exciton states) involved in the unique ground state indicates that the CFs near the pins become strongly interacting with new physics emerging beyond the simple mapping between the FQH of electrons and the IQH of the CFs. We conjecture that the parton description (which generalizes the CF description of the FQH states)~\cite{jain1989incompressible} may be needed to fully capture the fractionalization of the Laughlin quasiholes. 

\section{Considerations for Realistic Systems}

\begin{figure}
\includegraphics[width=\linewidth]{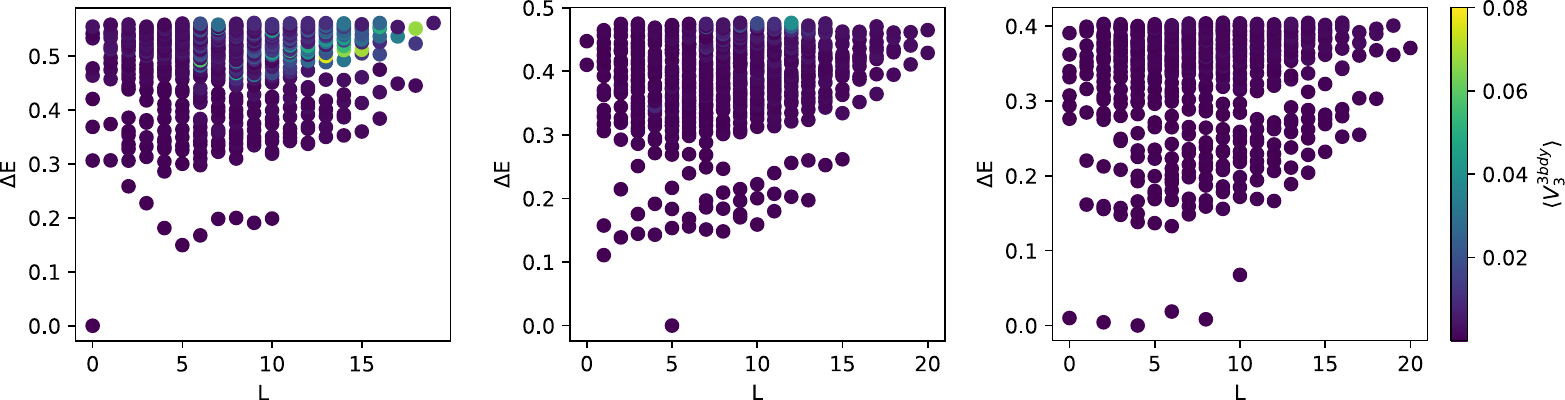}
\caption{The spectrum of Coulomb interaction projected to the LLL for systems with 10 electrons and 28, 29, and 30 orbitals (left to right). The color shows the $\hat V_3^{3bdy}$ variational energy of the corresponding eigenstate.}
\label{fig:Laughlin V3bdy Coulomb}
\end{figure}

We also consider the case where $\hat V^{2bdy}$ is not the $\hat V_1^{2bdy}$ pseudopotential but a more realistic Coulomb interaction (denoted as $\hat V_{\text{Coulomb}}$). Though a fully analytical proof remains elusive, it is strongly believed that $\hat V_1^{2bdy}$ is $\hat V_{\text{Coulomb}}$ are adiabatically connected. Furthermore, the ground states of these two interactions have extremely high overlap -- for a system with 10 electrons this overlap is above 0.99 for the ground state and one-quasihole state, and above 0.98 for the two-quasihole state. (Here, since multiple highest-weight states exist for two-quasihole, we take the cumulative overlap defined as the square root of the sum of squares of the individual overlaps.) Furthermore, just like in the case of $\hat V_1^{2bdy}$, the low-lying energy state of $\hat V_{\text{Coulomb}}$ is largely contained within $\mathcal H_{\text{MR}}$, as evidenced in the low $\hat V_3^{3bdy}$ variational energies of the eigenstates as shown in \Cref{fig:Laughlin V3bdy Coulomb}.

\begin{figure}
\includegraphics[width=\linewidth]{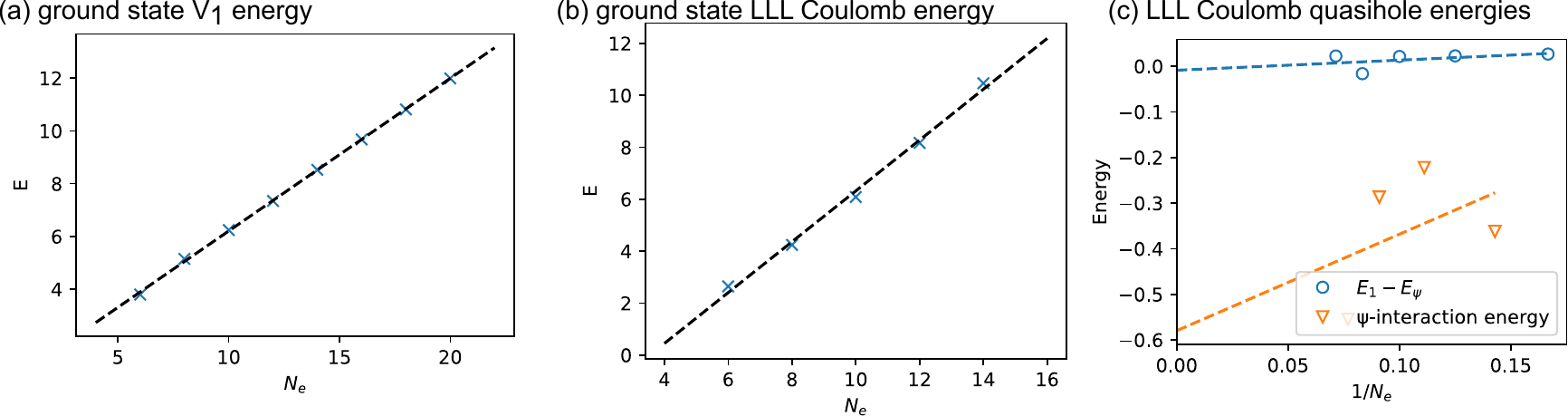}
\caption{(a-b) The variational energy of the MR ground state with respect to (a) $\hat V_1^{2bdy}$ and (b) LLL-projected Coulomb interaction. The dashed line shows a linear fit. (c) Difference between self-energy w.r.t. LLL Coulomb interaction of the $1$-quasihole and the $\psi$-quasihole (blue) and the average $\psi$-quasihole interaction energy at filling factor $\nu=1/3$ w.r.t to LLL Coulomb interaction. Dashed lines show linear fits.}
\label{fig:LLL Coulomb energy}
\end{figure}

Due to this adiabatic connection as well as the high ground state overlap, it is thus expected that the ground state of $\hat V_{\text{Coulomb}}$ at filling factor $\nu=1/3$ (which is closer to what is realized in experiment) can also be described as a fluid of $\psi$-anyon. However, here the dynamical origin of this picture is slightly different. Whereas under $\hat V_1^{2bdy}$ $\psi$-quasiholes are weakly interacting and energetically favoured (as shown in the main text), under Coulomb interaction $\psi$-quasiholes are strongly attractive (see \Cref{fig:LLL Coulomb energy}c). This strong attraction causes the creation of $\psi$-quasiholes to be energetically favoured at filling factor $\nu=1/3$, despite the fact that Coulomb interaction slightly prefers $1$-quasihole. A note of caution must be made here that finite-size scaling is hindered by computational limitation. Using second-quantized wavefuctions constructed from Jack polynomials \cite{bernevig2008model}, a maximum system size of 14 electrons is shown in \Cref{fig:LLL Coulomb energy}. Alternative computational tools, such as Monte-Carlo integration, may enable accessing much larger system sizes, but with some different caveats when it comes to analyzing anyon dynamics (e.g. ambiguity in constructing the appropriate first-quantized wavefunctions).  

\begin{figure}
\includegraphics[width=\linewidth]{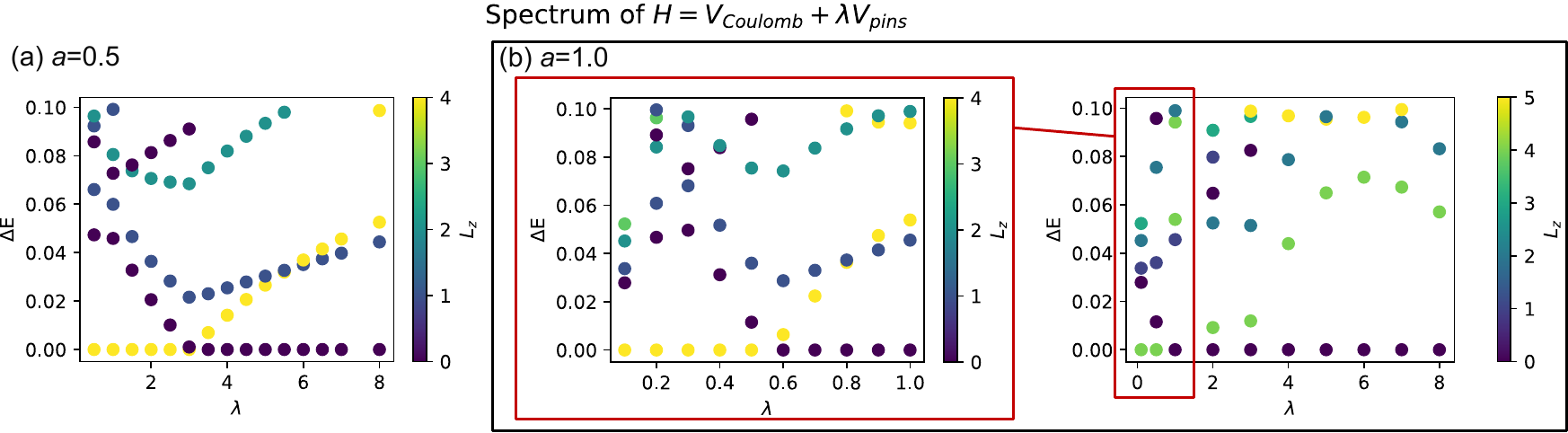}
\caption{Spectrum of \Cref{12body gen} where $\hat V^{2bdy}=\hat V_{\text{Coulomb}}$ is the Coulomb interaction projected to LLL level and $\hat V^{1bdy}$ consists of two Gaussian pins placed at two poles with (a) $a=0.5$, and (b) $a=1.0$ (at two different ranges of $\lambda$). All data are computed for a system with 8 electrons and 23 orbitals.}
\label{fig:coulomb and gaussian}
\end{figure}

Using the same angular momentum basis, projection of the Coulomb potential onto the LLL can be expressed in terms of the different pseudopotentials $\hat V_{m}^{2bdy}$ for $m=1,3,5,...$. Taking $\hat V^{1bdy}$ to be two Gaussian pins placed at the two poles, we again observe a ground state transition from $L_z=\pm N_e/2$ to $L_z=0$. For a fixed Gaussian width $a$, the transition occurs at the a smaller $\lambda$ compared to the $\hat V_1^{2bdy}$ case above. Most notably, the re-transition from $L_z=0$ to $L_z=\pm N_e/2$ is absent even at relatively large $\lambda$ ($\sim40$ times the Coulomb's roton gap). The fact that Laughlin quasihole fractionalization occurs more easily with Coulomb interaction compared to $\hat V_1^{2bdy}$ may be understood as a result of the terms $\hat V_3^{2bdy}$, $\hat V_5^{2bdy}$,... present in $\hat V_{\text{Coulomb}}$, which slightly favor fractionalized Laughlin quasiholes \cite{trung2021fractionalization}.

The fractionalized quasihole state under Coulomb interaction is also mostly contained within the Moore-Read CHS as shown in \Cref{table:overlap}. Note that while $\mathcal H_{\text{MR}}$ is a larger Hilbert space compared to $\mathcal H_{\text{L}}$ and hence a higher overlap is generally expected, its dimension is still very small compared to the entire LLL Hilbert space -- in the thermodynamic limit $\mathcal H_{\text{MR}}$ is still a subspace of measure zero compared to LLL. Thus, having such a high overlap is still an impressive and non-trivial result.

\begin{table}
\begin{tabular}{|c|c|c|c|c|}
\hline
$|\mathcal O|$& $\hat V_1^{2bdy}+$2 pins & $\hat V_{\text{Coulomb}}+$pins & $\mathcal H_{\text{L}}$ & $\mathcal H_{\text{MR}}$\\
\hline
$\hat V_1^{2bdy}+$2 pins &1& 0.890262958&0.0006730517&0.990902965\\
$\hat V_{\text{Coulomb}}+$2 pins &&1&0.0002645797&0.975764330\\
$\mathcal H_{\text{L}}$ &&&9&-\\
$\mathcal H_{\text{MR}}$ &&&-&80740\\
\hline
\end{tabular}
\caption{Comparison of the magnitude of the overlap between different ground states and CHS's, computed for a system with 8 electrons and 23 orbitals. From left to right on the first row the notations are: $\hat V_1^{2bdy}+$pins -- ground state of \Cref{12body gen} with $\hat V^{2bdy}=\hat V_1^{2bdy}$ and $\hat V^{1bdy}$ consisting of two pins at opposite pole, with $\lambda=2$, $\hat V_{\text{Coulomb}}+$2 pins -- ground state of \Cref{12body gen} with $\hat V^{2bdy}=\hat V_{\text{Coulomb}}$ and $\hat V^{1bdy}$ consisting of two pins at opposite pole, with $\lambda=2$, $\mathcal H_{\text{L}}$ -- the Laughlin CHS, and $\mathcal H_{\text{MR}}$ -- the Moore-Read CHS. The diagonal entries (``self-overlap'') gives the dimension of the corresponding subspace. For clarity, only upper-triangular terms are shown.}
\label{table:overlap}
\end{table}

\begin{figure}
\includegraphics[width=0.5\linewidth]{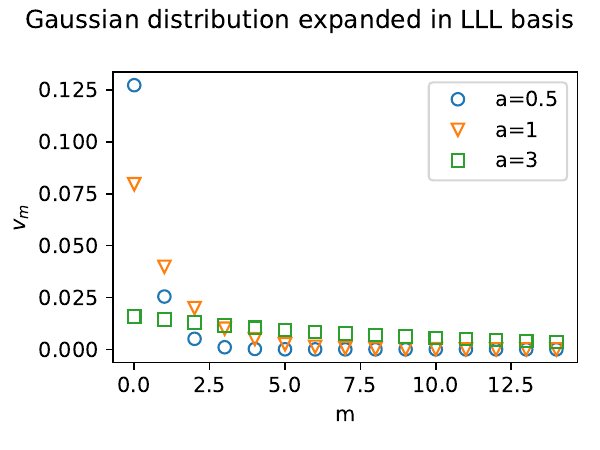}
\caption{The value of $v_m=\langle m|\hat V|m\rangle$ in the expansion in the LLL angular momentum basis of the Gaussian potential in \Cref{gaussian}.}
\label{fig:Gaussian coefficients}
\end{figure}

\subsection{Non-Abelian gap comparison}
Although there exist evidences that the filling factor $\nu=5/2$ can support the MR phase, to realize non-Abelian braiding is complicated by the realistic experimental conditions. This is due to, among other factors, the effect anyon-anyon interaction induced by two-body interactions, which is non-zero since the MR quasiholes are only exact zero-energy states of the three-body potential. A crucial quantity here is the true energy gap between the two non-Abelian states each containing four $\sigma$-quasiholes with fixed locations. These two states are exactly degenerate under the model interaction $\hat V_3^{3bdy}$, and also expected to be exactly degenerate under \emph{any} interaction in the thermodynamic limit provided each quasihole is infinitely far away from the others. In general, however, this energy gap suffers severely from finite-size effect, which is expected in real experiment on a small sample.

\begin{figure}
\includegraphics[width=0.5\linewidth]{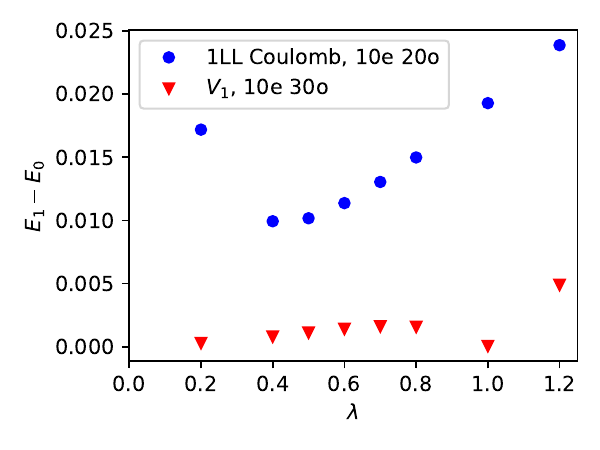}
\caption{The gap between the two lowest eigenstate with four potential pins applied to a system with two fluxes (the expected two-fold non-Abelian degeneracy), computed for Coulomb interaction at filling factor $\nu=1/2$ (blue) and $\hat V_1^{2bdy}$ interaction at filling factor $\nu=1/3$ (red). Both are shown for a system of 10 electrons.}
\label{fig:gap comparison}
\end{figure}

To investigate whether the filling factor $\nu=1/3$ can provide a more viable platform for non-Abelian braiding, we investigate this gap for a Hamiltonian of the form $\Cref{12body gen}$ where $\hat V^{1bdy}$ consists of four potential pins positioned at the vertices of a regular tetrahedron on the sphere. At the filling factor $\nu=5/2$, $\hat V^{2bdy}$ is the Coulomb interaction projected onto the $n=1$ Landau level (1LL). This Hamiltonian is diagonalized within a LLL Hilbert space satisfying the commensurability condition $N_o = 2N_e$. On the other hand, the results for filling factor $\nu=1/3$ is done by diagonalizing a Hamiltonian with $\hat V^{2bdy}=\hat V_1^{2bdy}$ at $N_o=3N_e$. The result for 10 electrons is shown in \Cref{fig:gap comparison}, which shows the non-Abelian gap is significantly smaller for $\nu=1/3$ (red triangles) compared to $\nu=1/2$ (blue circles). Thus, the filling factor $\nu=1/3$ not only shows a potential non-Abelian braiding, it is also an even more promising platform compared to the $\nu=5/2$ plateau that is currently the focus of intense research.

\bibliography{ref}

\begin{thebibliography}{87}%
\makeatletter
\providecommand \@ifxundefined [1]{%
 \@ifx{#1\undefined}
}%
\providecommand \@ifnum [1]{%
 \ifnum #1\expandafter \@firstoftwo
 \else \expandafter \@secondoftwo
 \fi
}%
\providecommand \@ifx [1]{%
 \ifx #1\expandafter \@firstoftwo
 \else \expandafter \@secondoftwo
 \fi
}%
\providecommand \natexlab [1]{#1}%
\providecommand \enquote  [1]{``#1''}%
\providecommand \bibnamefont  [1]{#1}%
\providecommand \bibfnamefont [1]{#1}%
\providecommand \citenamefont [1]{#1}%
\providecommand \href@noop [0]{\@secondoftwo}%
\providecommand \href [0]{\begingroup \@sanitize@url \@href}%
\providecommand \@href[1]{\@@startlink{#1}\@@href}%
\providecommand \@@href[1]{\endgroup#1\@@endlink}%
\providecommand \@sanitize@url [0]{\catcode `\\12\catcode `\$12\catcode
  `\&12\catcode `\#12\catcode `\^12\catcode `\_12\catcode `\%12\relax}%
\providecommand \@@startlink[1]{}%
\providecommand \@@endlink[0]{}%
\providecommand \url  [0]{\begingroup\@sanitize@url \@url }%
\providecommand \@url [1]{\endgroup\@href {#1}{\urlprefix }}%
\providecommand \urlprefix  [0]{URL }%
\providecommand \Eprint [0]{\href }%
\providecommand \doibase [0]{https://doi.org/}%
\providecommand \selectlanguage [0]{\@gobble}%
\providecommand \bibinfo  [0]{\@secondoftwo}%
\providecommand \bibfield  [0]{\@secondoftwo}%
\providecommand \translation [1]{[#1]}%
\providecommand \BibitemOpen [0]{}%
\providecommand \bibitemStop [0]{}%
\providecommand \bibitemNoStop [0]{.\EOS\space}%
\providecommand \EOS [0]{\spacefactor3000\relax}%
\providecommand \BibitemShut  [1]{\csname bibitem#1\endcsname}%
\let\auto@bib@innerbib\@empty
\bibitem [{\citenamefont {Leinaas}\ and\ \citenamefont
  {Myrheim}(1977)}]{Leinaas1977}%
  \BibitemOpen
  \bibfield  {author} {\bibinfo {author} {\bibfnamefont {J.~M.}\ \bibnamefont
  {Leinaas}}\ and\ \bibinfo {author} {\bibfnamefont {J.}~\bibnamefont
  {Myrheim}},\ }\href {https://doi.org/10.1007/BF02727953} {\bibinfo {title}
  {On the theory of identical particles}} (\bibinfo {year} {1977})\BibitemShut
  {NoStop}%
\bibitem [{\citenamefont {Wilczek}(1990)}]{wilczek1990fractional}%
  \BibitemOpen
  \bibfield  {author} {\bibinfo {author} {\bibfnamefont {F.}~\bibnamefont
  {Wilczek}},\ }\href@noop {} {\emph {\bibinfo {title} {Fractional statistics
  and anyon superconductivity}}},\ Vol.~\bibinfo {volume} {5}\ (\bibinfo
  {publisher} {World scientific},\ \bibinfo {year} {1990})\BibitemShut
  {NoStop}%
\bibitem [{\citenamefont {Simon}(2023)}]{simon2023topological}%
  \BibitemOpen
  \bibfield  {author} {\bibinfo {author} {\bibfnamefont {S.~H.}\ \bibnamefont
  {Simon}},\ }\href@noop {} {\emph {\bibinfo {title} {Topological quantum}}}\
  (\bibinfo  {publisher} {Oxford University Press},\ \bibinfo {year}
  {2023})\BibitemShut {NoStop}%
\bibitem [{\citenamefont {Nayak}\ \emph {et~al.}(2008)\citenamefont {Nayak},
  \citenamefont {Simon}, \citenamefont {Stern}, \citenamefont {Freedman},\ and\
  \citenamefont {Sarma}}]{Nayak2008}%
  \BibitemOpen
  \bibfield  {author} {\bibinfo {author} {\bibfnamefont {C.}~\bibnamefont
  {Nayak}}, \bibinfo {author} {\bibfnamefont {S.~H.}\ \bibnamefont {Simon}},
  \bibinfo {author} {\bibfnamefont {A.}~\bibnamefont {Stern}}, \bibinfo
  {author} {\bibfnamefont {M.}~\bibnamefont {Freedman}},\ and\ \bibinfo
  {author} {\bibfnamefont {S.~D.}\ \bibnamefont {Sarma}},\ }\bibfield  {title}
  {\bibinfo {title} {Non-abelian anyons and topological quantum computation},\
  }\href {https://doi.org/10.1103/RevModPhys.80.1083} {\bibfield  {journal}
  {\bibinfo  {journal} {Reviews of Modern Physics}\ }\textbf {\bibinfo {volume}
  {80}},\ \bibinfo {pages} {1083} (\bibinfo {year} {2008})}\BibitemShut
  {NoStop}%
\bibitem [{\citenamefont {Freedman}\ \emph {et~al.}(2000)\citenamefont
  {Freedman}, \citenamefont {Larsen},\ and\ \citenamefont
  {Wang}}]{Freedman2000}%
  \BibitemOpen
  \bibfield  {author} {\bibinfo {author} {\bibfnamefont {M.}~\bibnamefont
  {Freedman}}, \bibinfo {author} {\bibfnamefont {M.}~\bibnamefont {Larsen}},\
  and\ \bibinfo {author} {\bibfnamefont {Z.}~\bibnamefont {Wang}},\ }\href@noop
  {} {\bibinfo {title} {A modular functor which is universal for quantum
  computation}} (\bibinfo {year} {2000})\BibitemShut {NoStop}%
\bibitem [{\citenamefont {Sarma}\ \emph {et~al.}(2015)\citenamefont {Sarma},
  \citenamefont {Freedman},\ and\ \citenamefont {Nayak}}]{sarma2015majorana}%
  \BibitemOpen
  \bibfield  {author} {\bibinfo {author} {\bibfnamefont {S.~D.}\ \bibnamefont
  {Sarma}}, \bibinfo {author} {\bibfnamefont {M.}~\bibnamefont {Freedman}},\
  and\ \bibinfo {author} {\bibfnamefont {C.}~\bibnamefont {Nayak}},\ }\bibfield
   {title} {\bibinfo {title} {Majorana zero modes and topological quantum
  computation},\ }\href@noop {} {\bibfield  {journal} {\bibinfo  {journal} {npj
  Quantum Information}\ }\textbf {\bibinfo {volume} {1}},\ \bibinfo {pages} {1}
  (\bibinfo {year} {2015})}\BibitemShut {NoStop}%
\bibitem [{\citenamefont {Kitaev}(2003)}]{kitaev2003fault}%
  \BibitemOpen
  \bibfield  {author} {\bibinfo {author} {\bibfnamefont {A.~Y.}\ \bibnamefont
  {Kitaev}},\ }\bibfield  {title} {\bibinfo {title} {Fault-tolerant quantum
  computation by anyons},\ }\href@noop {} {\bibfield  {journal} {\bibinfo
  {journal} {Annals of physics}\ }\textbf {\bibinfo {volume} {303}},\ \bibinfo
  {pages} {2} (\bibinfo {year} {2003})}\BibitemShut {NoStop}%
\bibitem [{\citenamefont {Das~Sarma}\ \emph {et~al.}(2005)\citenamefont
  {Das~Sarma}, \citenamefont {Freedman},\ and\ \citenamefont
  {Nayak}}]{das2005topologically}%
  \BibitemOpen
  \bibfield  {author} {\bibinfo {author} {\bibfnamefont {S.}~\bibnamefont
  {Das~Sarma}}, \bibinfo {author} {\bibfnamefont {M.}~\bibnamefont
  {Freedman}},\ and\ \bibinfo {author} {\bibfnamefont {C.}~\bibnamefont
  {Nayak}},\ }\bibfield  {title} {\bibinfo {title} {Topologically protected
  qubits from a possible non-abelian fractional quantum hall state},\
  }\href@noop {} {\bibfield  {journal} {\bibinfo  {journal} {Physical review
  letters}\ }\textbf {\bibinfo {volume} {94}},\ \bibinfo {pages} {166802}
  (\bibinfo {year} {2005})}\BibitemShut {NoStop}%
\bibitem [{\citenamefont {Tsui}\ \emph {et~al.}(1982)\citenamefont {Tsui},
  \citenamefont {Stormer},\ and\ \citenamefont {Gossard}}]{tsui1982two}%
  \BibitemOpen
  \bibfield  {author} {\bibinfo {author} {\bibfnamefont {D.~C.}\ \bibnamefont
  {Tsui}}, \bibinfo {author} {\bibfnamefont {H.~L.}\ \bibnamefont {Stormer}},\
  and\ \bibinfo {author} {\bibfnamefont {A.~C.}\ \bibnamefont {Gossard}},\
  }\bibfield  {title} {\bibinfo {title} {Two-dimensional magnetotransport in
  the extreme quantum limit},\ }\href@noop {} {\bibfield  {journal} {\bibinfo
  {journal} {Physical Review Letters}\ }\textbf {\bibinfo {volume} {48}},\
  \bibinfo {pages} {1559} (\bibinfo {year} {1982})}\BibitemShut {NoStop}%
\bibitem [{\citenamefont {Yoshioka}\ \emph {et~al.}(1984)\citenamefont
  {Yoshioka}, \citenamefont {Halperin},\ and\ \citenamefont
  {Lee}}]{yoshioka1984ground}%
  \BibitemOpen
  \bibfield  {author} {\bibinfo {author} {\bibfnamefont {D.}~\bibnamefont
  {Yoshioka}}, \bibinfo {author} {\bibfnamefont {B.}~\bibnamefont {Halperin}},\
  and\ \bibinfo {author} {\bibfnamefont {P.}~\bibnamefont {Lee}},\ }\bibfield
  {title} {\bibinfo {title} {The ground state of the 2d electrons in a strong
  magnetic field and the anomalous quantized hall effect},\ }\href@noop {}
  {\bibfield  {journal} {\bibinfo  {journal} {Surface Science}\ }\textbf
  {\bibinfo {volume} {142}},\ \bibinfo {pages} {155} (\bibinfo {year}
  {1984})}\BibitemShut {NoStop}%
\bibitem [{\citenamefont {Arovas}\ \emph {et~al.}(1984)\citenamefont {Arovas},
  \citenamefont {Schrieffer},\ and\ \citenamefont
  {Wilczek}}]{arovas1984fractional}%
  \BibitemOpen
  \bibfield  {author} {\bibinfo {author} {\bibfnamefont {D.}~\bibnamefont
  {Arovas}}, \bibinfo {author} {\bibfnamefont {J.~R.}\ \bibnamefont
  {Schrieffer}},\ and\ \bibinfo {author} {\bibfnamefont {F.}~\bibnamefont
  {Wilczek}},\ }\bibfield  {title} {\bibinfo {title} {Fractional statistics and
  the quantum hall effect},\ }\href@noop {} {\bibfield  {journal} {\bibinfo
  {journal} {Physical review letters}\ }\textbf {\bibinfo {volume} {53}},\
  \bibinfo {pages} {722} (\bibinfo {year} {1984})}\BibitemShut {NoStop}%
\bibitem [{\citenamefont {Moore}\ and\ \citenamefont
  {Read}(1991)}]{moore1991nonabelions}%
  \BibitemOpen
  \bibfield  {author} {\bibinfo {author} {\bibfnamefont {G.}~\bibnamefont
  {Moore}}\ and\ \bibinfo {author} {\bibfnamefont {N.}~\bibnamefont {Read}},\
  }\bibfield  {title} {\bibinfo {title} {Nonabelions in the fractional quantum
  hall effect},\ }\href@noop {} {\bibfield  {journal} {\bibinfo  {journal}
  {Nuclear Physics B}\ }\textbf {\bibinfo {volume} {360}},\ \bibinfo {pages}
  {362} (\bibinfo {year} {1991})}\BibitemShut {NoStop}%
\bibitem [{\citenamefont {Nayak}\ and\ \citenamefont
  {Wilczek}(1996)}]{nayak19962n}%
  \BibitemOpen
  \bibfield  {author} {\bibinfo {author} {\bibfnamefont {C.}~\bibnamefont
  {Nayak}}\ and\ \bibinfo {author} {\bibfnamefont {F.}~\bibnamefont
  {Wilczek}},\ }\bibfield  {title} {\bibinfo {title} {2n-quasihole states
  realize 2n- 1-dimensional spinor braiding statistics in paired quantum hall
  states},\ }\href@noop {} {\bibfield  {journal} {\bibinfo  {journal} {Nuclear
  Physics B}\ }\textbf {\bibinfo {volume} {479}},\ \bibinfo {pages} {529}
  (\bibinfo {year} {1996})}\BibitemShut {NoStop}%
\bibitem [{\citenamefont {Read}\ and\ \citenamefont {Rezayi}(1996)}]{Read1996}%
  \BibitemOpen
  \bibfield  {author} {\bibinfo {author} {\bibfnamefont {N.}~\bibnamefont
  {Read}}\ and\ \bibinfo {author} {\bibfnamefont {E.}~\bibnamefont {Rezayi}},\
  }\bibfield  {title} {\bibinfo {title} {Quasiholes and fermionic zero modes of
  paired fractional quantum hall states: The mechanism for non-abelian
  statistics},\ }\href {https://doi.org/10.1103/PhysRevB.54.16864} {\bibfield
  {journal} {\bibinfo  {journal} {Physical Review B - Condensed Matter and
  Materials Physics}\ }\textbf {\bibinfo {volume} {54}},\ \bibinfo {pages}
  {16864} (\bibinfo {year} {1996})}\BibitemShut {NoStop}%
\bibitem [{\citenamefont {Feldman}\ and\ \citenamefont
  {Halperin}(2021)}]{feldman2021}%
  \BibitemOpen
  \bibfield  {author} {\bibinfo {author} {\bibfnamefont {D.~E.}\ \bibnamefont
  {Feldman}}\ and\ \bibinfo {author} {\bibfnamefont {B.~I.}\ \bibnamefont
  {Halperin}},\ }\bibfield  {title} {\bibinfo {title} {Fractional charge and
  fractional statistics in the quantum hall effects},\ }\href
  {http://arxiv.org/abs/2102.08998} {\ ,\ \bibinfo {pages} {1} (\bibinfo {year}
  {2021})}\BibitemShut {NoStop}%
\bibitem [{\citenamefont {Kitaev}(2001)}]{kitaev2001unpaired}%
  \BibitemOpen
  \bibfield  {author} {\bibinfo {author} {\bibfnamefont {A.~Y.}\ \bibnamefont
  {Kitaev}},\ }\bibfield  {title} {\bibinfo {title} {Unpaired majorana fermions
  in quantum wires},\ }\href@noop {} {\bibfield  {journal} {\bibinfo  {journal}
  {Physics-uspekhi}\ }\textbf {\bibinfo {volume} {44}},\ \bibinfo {pages} {131}
  (\bibinfo {year} {2001})}\BibitemShut {NoStop}%
\bibitem [{\citenamefont {Alicea}(2012)}]{alicea2012new}%
  \BibitemOpen
  \bibfield  {author} {\bibinfo {author} {\bibfnamefont {J.}~\bibnamefont
  {Alicea}},\ }\bibfield  {title} {\bibinfo {title} {New directions in the
  pursuit of majorana fermions in solid state systems},\ }\href@noop {}
  {\bibfield  {journal} {\bibinfo  {journal} {Reports on progress in physics}\
  }\textbf {\bibinfo {volume} {75}},\ \bibinfo {pages} {076501} (\bibinfo
  {year} {2012})}\BibitemShut {NoStop}%
\bibitem [{\citenamefont {Hu}\ \emph {et~al.}(2015)\citenamefont {Hu},
  \citenamefont {Cai}, \citenamefont {Baranov},\ and\ \citenamefont
  {Zoller}}]{hu2015majorana}%
  \BibitemOpen
  \bibfield  {author} {\bibinfo {author} {\bibfnamefont {Y.}~\bibnamefont
  {Hu}}, \bibinfo {author} {\bibfnamefont {Z.}~\bibnamefont {Cai}}, \bibinfo
  {author} {\bibfnamefont {M.~A.}\ \bibnamefont {Baranov}},\ and\ \bibinfo
  {author} {\bibfnamefont {P.}~\bibnamefont {Zoller}},\ }\bibfield  {title}
  {\bibinfo {title} {Majorana fermions in noisy kitaev wires},\ }\href@noop {}
  {\bibfield  {journal} {\bibinfo  {journal} {Physical Review B}\ }\textbf
  {\bibinfo {volume} {92}},\ \bibinfo {pages} {165118} (\bibinfo {year}
  {2015})}\BibitemShut {NoStop}%
\bibitem [{\citenamefont {Franz}(2010)}]{franz2010race}%
  \BibitemOpen
  \bibfield  {author} {\bibinfo {author} {\bibfnamefont {M.}~\bibnamefont
  {Franz}},\ }\bibfield  {title} {\bibinfo {title} {Race for majorana
  fermions},\ }\href@noop {} {\bibfield  {journal} {\bibinfo  {journal}
  {Physics}\ }\textbf {\bibinfo {volume} {3}},\ \bibinfo {pages} {24} (\bibinfo
  {year} {2010})}\BibitemShut {NoStop}%
\bibitem [{\citenamefont {Kitaev}(2006)}]{kitaev2006anyons}%
  \BibitemOpen
  \bibfield  {author} {\bibinfo {author} {\bibfnamefont {A.}~\bibnamefont
  {Kitaev}},\ }\bibfield  {title} {\bibinfo {title} {Anyons in an exactly
  solved model and beyond},\ }\href@noop {} {\bibfield  {journal} {\bibinfo
  {journal} {Annals of Physics}\ }\textbf {\bibinfo {volume} {321}},\ \bibinfo
  {pages} {2} (\bibinfo {year} {2006})}\BibitemShut {NoStop}%
\bibitem [{\citenamefont {Tsintzis}\ \emph {et~al.}(2024)\citenamefont
  {Tsintzis}, \citenamefont {Souto}, \citenamefont {Flensberg}, \citenamefont
  {Danon},\ and\ \citenamefont {Leijnse}}]{tsintzis2024majorana}%
  \BibitemOpen
  \bibfield  {author} {\bibinfo {author} {\bibfnamefont {A.}~\bibnamefont
  {Tsintzis}}, \bibinfo {author} {\bibfnamefont {R.~S.}\ \bibnamefont {Souto}},
  \bibinfo {author} {\bibfnamefont {K.}~\bibnamefont {Flensberg}}, \bibinfo
  {author} {\bibfnamefont {J.}~\bibnamefont {Danon}},\ and\ \bibinfo {author}
  {\bibfnamefont {M.}~\bibnamefont {Leijnse}},\ }\bibfield  {title} {\bibinfo
  {title} {Majorana qubits and non-abelian physics in quantum dot--based
  minimal kitaev chains},\ }\href@noop {} {\bibfield  {journal} {\bibinfo
  {journal} {PRX quantum}\ }\textbf {\bibinfo {volume} {5}},\ \bibinfo {pages}
  {010323} (\bibinfo {year} {2024})}\BibitemShut {NoStop}%
\bibitem [{\citenamefont {Li}\ \emph {et~al.}(2014)\citenamefont {Li},
  \citenamefont {Li}, \citenamefont {Cai},\ and\ \citenamefont
  {Sun}}]{li2014probing}%
  \BibitemOpen
  \bibfield  {author} {\bibinfo {author} {\bibfnamefont {S.-W.}\ \bibnamefont
  {Li}}, \bibinfo {author} {\bibfnamefont {Z.-Z.}\ \bibnamefont {Li}}, \bibinfo
  {author} {\bibfnamefont {C.}~\bibnamefont {Cai}},\ and\ \bibinfo {author}
  {\bibfnamefont {C.}~\bibnamefont {Sun}},\ }\bibfield  {title} {\bibinfo
  {title} {Probing zero modes of a defect in a kitaev quantum wire},\
  }\href@noop {} {\bibfield  {journal} {\bibinfo  {journal} {Physical Review
  B}\ }\textbf {\bibinfo {volume} {89}},\ \bibinfo {pages} {134505} (\bibinfo
  {year} {2014})}\BibitemShut {NoStop}%
\bibitem [{goo(2023)}]{google2023non}%
  \BibitemOpen
  \bibfield  {title} {\bibinfo {title} {Non-abelian braiding of graph vertices
  in a superconducting processor},\ }\href@noop {} {\bibfield  {journal}
  {\bibinfo  {journal} {Nature}\ }\textbf {\bibinfo {volume} {618}},\ \bibinfo
  {pages} {264} (\bibinfo {year} {2023})}\BibitemShut {NoStop}%
\bibitem [{\citenamefont {Aghaee}\ \emph {et~al.}(2023)\citenamefont {Aghaee},
  \citenamefont {Akkala}, \citenamefont {Alam}, \citenamefont {Ali},
  \citenamefont {Alcaraz~Ramirez}, \citenamefont {Andrzejczuk}, \citenamefont
  {Antipov}, \citenamefont {Aseev}, \citenamefont {Astafev}, \citenamefont
  {Bauer} \emph {et~al.}}]{aghaee2023inas}%
  \BibitemOpen
  \bibfield  {author} {\bibinfo {author} {\bibfnamefont {M.}~\bibnamefont
  {Aghaee}}, \bibinfo {author} {\bibfnamefont {A.}~\bibnamefont {Akkala}},
  \bibinfo {author} {\bibfnamefont {Z.}~\bibnamefont {Alam}}, \bibinfo {author}
  {\bibfnamefont {R.}~\bibnamefont {Ali}}, \bibinfo {author} {\bibfnamefont
  {A.}~\bibnamefont {Alcaraz~Ramirez}}, \bibinfo {author} {\bibfnamefont
  {M.}~\bibnamefont {Andrzejczuk}}, \bibinfo {author} {\bibfnamefont {A.~E.}\
  \bibnamefont {Antipov}}, \bibinfo {author} {\bibfnamefont {P.}~\bibnamefont
  {Aseev}}, \bibinfo {author} {\bibfnamefont {M.}~\bibnamefont {Astafev}},
  \bibinfo {author} {\bibfnamefont {B.}~\bibnamefont {Bauer}}, \emph {et~al.},\
  }\bibfield  {title} {\bibinfo {title} {Inas-al hybrid devices passing the
  topological gap protocol},\ }\href@noop {} {\bibfield  {journal} {\bibinfo
  {journal} {Physical Review B}\ }\textbf {\bibinfo {volume} {107}},\ \bibinfo
  {pages} {245423} (\bibinfo {year} {2023})}\BibitemShut {NoStop}%
\bibitem [{\citenamefont {Quantum}\ \emph {et~al.}(2025)\citenamefont
  {Quantum}, \citenamefont {Aghaee}, \citenamefont {Alcaraz~Ramirez},
  \citenamefont {Alam}, \citenamefont {Ali}, \citenamefont {Andrzejczuk},
  \citenamefont {Antipov}, \citenamefont {Astafev}, \citenamefont {Barzegar},
  \citenamefont {Bauer} \emph {et~al.}}]{microsoft2025interferometric}%
  \BibitemOpen
  \bibfield  {author} {\bibinfo {author} {\bibfnamefont {M.~A.}\ \bibnamefont
  {Quantum}}, \bibinfo {author} {\bibfnamefont {M.}~\bibnamefont {Aghaee}},
  \bibinfo {author} {\bibfnamefont {A.}~\bibnamefont {Alcaraz~Ramirez}},
  \bibinfo {author} {\bibfnamefont {Z.}~\bibnamefont {Alam}}, \bibinfo {author}
  {\bibfnamefont {R.}~\bibnamefont {Ali}}, \bibinfo {author} {\bibfnamefont
  {M.}~\bibnamefont {Andrzejczuk}}, \bibinfo {author} {\bibfnamefont
  {A.}~\bibnamefont {Antipov}}, \bibinfo {author} {\bibfnamefont
  {M.}~\bibnamefont {Astafev}}, \bibinfo {author} {\bibfnamefont
  {A.}~\bibnamefont {Barzegar}}, \bibinfo {author} {\bibfnamefont
  {B.}~\bibnamefont {Bauer}}, \emph {et~al.},\ }\bibfield  {title} {\bibinfo
  {title} {Interferometric single-shot parity measurement in inas--al hybrid
  devices},\ }\href@noop {} {\bibfield  {journal} {\bibinfo  {journal}
  {Nature}\ }\textbf {\bibinfo {volume} {638}},\ \bibinfo {pages} {651}
  (\bibinfo {year} {2025})}\BibitemShut {NoStop}%
\bibitem [{\citenamefont {Aasen}\ \emph {et~al.}(2025)\citenamefont {Aasen},
  \citenamefont {Aghaee}, \citenamefont {Alam}, \citenamefont {Andrzejczuk},
  \citenamefont {Antipov}, \citenamefont {Astafev}, \citenamefont {Avilovas},
  \citenamefont {Barzegar}, \citenamefont {Bauer}, \citenamefont {Becker} \emph
  {et~al.}}]{aasen2025roadmap}%
  \BibitemOpen
  \bibfield  {author} {\bibinfo {author} {\bibfnamefont {D.}~\bibnamefont
  {Aasen}}, \bibinfo {author} {\bibfnamefont {M.}~\bibnamefont {Aghaee}},
  \bibinfo {author} {\bibfnamefont {Z.}~\bibnamefont {Alam}}, \bibinfo {author}
  {\bibfnamefont {M.}~\bibnamefont {Andrzejczuk}}, \bibinfo {author}
  {\bibfnamefont {A.}~\bibnamefont {Antipov}}, \bibinfo {author} {\bibfnamefont
  {M.}~\bibnamefont {Astafev}}, \bibinfo {author} {\bibfnamefont
  {L.}~\bibnamefont {Avilovas}}, \bibinfo {author} {\bibfnamefont
  {A.}~\bibnamefont {Barzegar}}, \bibinfo {author} {\bibfnamefont
  {B.}~\bibnamefont {Bauer}}, \bibinfo {author} {\bibfnamefont
  {J.}~\bibnamefont {Becker}}, \emph {et~al.},\ }\bibfield  {title} {\bibinfo
  {title} {Roadmap to fault tolerant quantum computation using topological
  qubit arrays},\ }\href@noop {} {\bibfield  {journal} {\bibinfo  {journal}
  {arXiv preprint arXiv:2502.12252}\ } (\bibinfo {year} {2025})}\BibitemShut
  {NoStop}%
\bibitem [{\citenamefont {Frolov}(2021)}]{frolov2021quantum}%
  \BibitemOpen
  \bibfield  {author} {\bibinfo {author} {\bibfnamefont {S.}~\bibnamefont
  {Frolov}},\ }\bibfield  {title} {\bibinfo {title} {Quantum computing’s
  reproducibility crisis: Majorana fermions},\ }\href@noop {} {\bibfield
  {journal} {\bibinfo  {journal} {Nature}\ }\textbf {\bibinfo {volume} {592}},\
  \bibinfo {pages} {350} (\bibinfo {year} {2021})}\BibitemShut {NoStop}%
\bibitem [{\citenamefont {Rini}(2025)}]{rini2025microsoft}%
  \BibitemOpen
  \bibfield  {author} {\bibinfo {author} {\bibfnamefont {M.}~\bibnamefont
  {Rini}},\ }\bibfield  {title} {\bibinfo {title} {Microsoft’s claim of a
  topological qubit faces tough questions},\ }\href@noop {} {\bibfield
  {journal} {\bibinfo  {journal} {Physics}\ }\textbf {\bibinfo {volume} {18}},\
  \bibinfo {pages} {68} (\bibinfo {year} {2025})}\BibitemShut {NoStop}%
\bibitem [{\citenamefont {Willett}\ \emph {et~al.}(1987)\citenamefont
  {Willett}, \citenamefont {Eisenstein}, \citenamefont {St{\"o}rmer},
  \citenamefont {Tsui}, \citenamefont {Gossard},\ and\ \citenamefont
  {English}}]{willett1987observation}%
  \BibitemOpen
  \bibfield  {author} {\bibinfo {author} {\bibfnamefont {R.}~\bibnamefont
  {Willett}}, \bibinfo {author} {\bibfnamefont {J.~P.}\ \bibnamefont
  {Eisenstein}}, \bibinfo {author} {\bibfnamefont {H.~L.}\ \bibnamefont
  {St{\"o}rmer}}, \bibinfo {author} {\bibfnamefont {D.~C.}\ \bibnamefont
  {Tsui}}, \bibinfo {author} {\bibfnamefont {A.~C.}\ \bibnamefont {Gossard}},\
  and\ \bibinfo {author} {\bibfnamefont {J.}~\bibnamefont {English}},\
  }\bibfield  {title} {\bibinfo {title} {Observation of an even-denominator
  quantum number in the fractional quantum hall effect},\ }\href@noop {}
  {\bibfield  {journal} {\bibinfo  {journal} {Physical review letters}\
  }\textbf {\bibinfo {volume} {59}},\ \bibinfo {pages} {1776} (\bibinfo {year}
  {1987})}\BibitemShut {NoStop}%
\bibitem [{\citenamefont {Pan}\ \emph {et~al.}(1999)\citenamefont {Pan},
  \citenamefont {Xia}, \citenamefont {Shvarts}, \citenamefont {Adams},
  \citenamefont {Stormer}, \citenamefont {Tsui}, \citenamefont {Pfeiffer},
  \citenamefont {Baldwin},\ and\ \citenamefont {West}}]{pan1999exact}%
  \BibitemOpen
  \bibfield  {author} {\bibinfo {author} {\bibfnamefont {W.}~\bibnamefont
  {Pan}}, \bibinfo {author} {\bibfnamefont {J.-S.}\ \bibnamefont {Xia}},
  \bibinfo {author} {\bibfnamefont {V.}~\bibnamefont {Shvarts}}, \bibinfo
  {author} {\bibfnamefont {D.}~\bibnamefont {Adams}}, \bibinfo {author}
  {\bibfnamefont {H.}~\bibnamefont {Stormer}}, \bibinfo {author} {\bibfnamefont
  {D.}~\bibnamefont {Tsui}}, \bibinfo {author} {\bibfnamefont {L.}~\bibnamefont
  {Pfeiffer}}, \bibinfo {author} {\bibfnamefont {K.}~\bibnamefont {Baldwin}},\
  and\ \bibinfo {author} {\bibfnamefont {K.}~\bibnamefont {West}},\ }\bibfield
  {title} {\bibinfo {title} {Exact quantization of the even-denominator
  fractional quantum hall state at nu = 5/2 landau level filling factor},\
  }\href@noop {} {\bibfield  {journal} {\bibinfo  {journal} {Physical review
  letters}\ }\textbf {\bibinfo {volume} {83}},\ \bibinfo {pages} {3530}
  (\bibinfo {year} {1999})}\BibitemShut {NoStop}%
\bibitem [{\citenamefont {Willett}\ \emph {et~al.}(2023)\citenamefont
  {Willett}, \citenamefont {Shtengel}, \citenamefont {Nayak}, \citenamefont
  {Pfeiffer}, \citenamefont {Chung}, \citenamefont {Peabody}, \citenamefont
  {Baldwin},\ and\ \citenamefont {West}}]{willett2023interference}%
  \BibitemOpen
  \bibfield  {author} {\bibinfo {author} {\bibfnamefont {R.}~\bibnamefont
  {Willett}}, \bibinfo {author} {\bibfnamefont {K.}~\bibnamefont {Shtengel}},
  \bibinfo {author} {\bibfnamefont {C.}~\bibnamefont {Nayak}}, \bibinfo
  {author} {\bibfnamefont {L.}~\bibnamefont {Pfeiffer}}, \bibinfo {author}
  {\bibfnamefont {Y.}~\bibnamefont {Chung}}, \bibinfo {author} {\bibfnamefont
  {M.}~\bibnamefont {Peabody}}, \bibinfo {author} {\bibfnamefont
  {K.}~\bibnamefont {Baldwin}},\ and\ \bibinfo {author} {\bibfnamefont
  {K.}~\bibnamefont {West}},\ }\bibfield  {title} {\bibinfo {title}
  {Interference measurements of non-abelian e/4 \& abelian e/2 quasiparticle
  braiding},\ }\href@noop {} {\bibfield  {journal} {\bibinfo  {journal}
  {Physical Review X}\ }\textbf {\bibinfo {volume} {13}},\ \bibinfo {pages}
  {011028} (\bibinfo {year} {2023})}\BibitemShut {NoStop}%
\bibitem [{\citenamefont {Levin}\ \emph {et~al.}(2007)\citenamefont {Levin},
  \citenamefont {Halperin},\ and\ \citenamefont {Rosenow}}]{levin2007particle}%
  \BibitemOpen
  \bibfield  {author} {\bibinfo {author} {\bibfnamefont {M.}~\bibnamefont
  {Levin}}, \bibinfo {author} {\bibfnamefont {B.~I.}\ \bibnamefont
  {Halperin}},\ and\ \bibinfo {author} {\bibfnamefont {B.}~\bibnamefont
  {Rosenow}},\ }\bibfield  {title} {\bibinfo {title} {Particle-hole symmetry
  and the pfaffian state},\ }\href@noop {} {\bibfield  {journal} {\bibinfo
  {journal} {Physical review letters}\ }\textbf {\bibinfo {volume} {99}},\
  \bibinfo {pages} {236806} (\bibinfo {year} {2007})}\BibitemShut {NoStop}%
\bibitem [{\citenamefont {Wang}\ \emph {et~al.}(2018)\citenamefont {Wang},
  \citenamefont {Vishwanath},\ and\ \citenamefont
  {Halperin}}]{wang2018topological}%
  \BibitemOpen
  \bibfield  {author} {\bibinfo {author} {\bibfnamefont {C.}~\bibnamefont
  {Wang}}, \bibinfo {author} {\bibfnamefont {A.}~\bibnamefont {Vishwanath}},\
  and\ \bibinfo {author} {\bibfnamefont {B.~I.}\ \bibnamefont {Halperin}},\
  }\bibfield  {title} {\bibinfo {title} {Topological order from disorder and
  the quantized hall thermal metal: Possible applications to the nu= 5/2
  state},\ }\href@noop {} {\bibfield  {journal} {\bibinfo  {journal} {Physical
  Review B}\ }\textbf {\bibinfo {volume} {98}},\ \bibinfo {pages} {045112}
  (\bibinfo {year} {2018})}\BibitemShut {NoStop}%
\bibitem [{\citenamefont {Ma}\ \emph {et~al.}(2022)\citenamefont {Ma},
  \citenamefont {Peterson}, \citenamefont {Scarola},\ and\ \citenamefont
  {Yang}}]{ma2022fractional}%
  \BibitemOpen
  \bibfield  {author} {\bibinfo {author} {\bibfnamefont {K.~K.}\ \bibnamefont
  {Ma}}, \bibinfo {author} {\bibfnamefont {M.~R.}\ \bibnamefont {Peterson}},
  \bibinfo {author} {\bibfnamefont {V.~W.}\ \bibnamefont {Scarola}},\ and\
  \bibinfo {author} {\bibfnamefont {K.}~\bibnamefont {Yang}},\ }\bibfield
  {title} {\bibinfo {title} {Fractional quantum hall effect at the filling
  factor nu= 5/2},\ }\href@noop {} {\bibfield  {journal} {\bibinfo  {journal}
  {arXiv preprint arXiv:2208.07908}\ } (\bibinfo {year} {2022})}\BibitemShut
  {NoStop}%
\bibitem [{\citenamefont {Asasi}\ and\ \citenamefont
  {Mulligan}(2020)}]{asasi2020partial}%
  \BibitemOpen
  \bibfield  {author} {\bibinfo {author} {\bibfnamefont {H.}~\bibnamefont
  {Asasi}}\ and\ \bibinfo {author} {\bibfnamefont {M.}~\bibnamefont
  {Mulligan}},\ }\bibfield  {title} {\bibinfo {title} {Partial equilibration of
  anti-pfaffian edge modes at nu= 5/2},\ }\href@noop {} {\bibfield  {journal}
  {\bibinfo  {journal} {arXiv preprint arXiv:2004.04161}\ } (\bibinfo {year}
  {2020})}\BibitemShut {NoStop}%
\bibitem [{\citenamefont {Hein}\ and\ \citenamefont
  {Sp{\aa}nsl{\"a}tt}(2023)}]{hein2023thermal}%
  \BibitemOpen
  \bibfield  {author} {\bibinfo {author} {\bibfnamefont {M.}~\bibnamefont
  {Hein}}\ and\ \bibinfo {author} {\bibfnamefont {C.}~\bibnamefont
  {Sp{\aa}nsl{\"a}tt}},\ }\bibfield  {title} {\bibinfo {title} {Thermal
  conductance and noise of majorana modes along interfaced nu = 5 2 fractional
  quantum hall states},\ }\href@noop {} {\bibfield  {journal} {\bibinfo
  {journal} {Physical Review B}\ }\textbf {\bibinfo {volume} {107}},\ \bibinfo
  {pages} {245301} (\bibinfo {year} {2023})}\BibitemShut {NoStop}%
\bibitem [{\citenamefont {Manna}\ \emph {et~al.}(2024)\citenamefont {Manna},
  \citenamefont {Das}, \citenamefont {Goldstein},\ and\ \citenamefont
  {Gefen}}]{manna2024full}%
  \BibitemOpen
  \bibfield  {author} {\bibinfo {author} {\bibfnamefont {S.}~\bibnamefont
  {Manna}}, \bibinfo {author} {\bibfnamefont {A.}~\bibnamefont {Das}}, \bibinfo
  {author} {\bibfnamefont {M.}~\bibnamefont {Goldstein}},\ and\ \bibinfo
  {author} {\bibfnamefont {Y.}~\bibnamefont {Gefen}},\ }\bibfield  {title}
  {\bibinfo {title} {Full classification of transport on an equilibrated 5/2
  edge via shot noise},\ }\href@noop {} {\bibfield  {journal} {\bibinfo
  {journal} {Physical Review Letters}\ }\textbf {\bibinfo {volume} {132}},\
  \bibinfo {pages} {136502} (\bibinfo {year} {2024})}\BibitemShut {NoStop}%
\bibitem [{\citenamefont {Dolev}\ \emph {et~al.}(2008)\citenamefont {Dolev},
  \citenamefont {Heiblum}, \citenamefont {Umansky}, \citenamefont {Stern},\
  and\ \citenamefont {Mahalu}}]{dolev2008observation}%
  \BibitemOpen
  \bibfield  {author} {\bibinfo {author} {\bibfnamefont {M.}~\bibnamefont
  {Dolev}}, \bibinfo {author} {\bibfnamefont {M.}~\bibnamefont {Heiblum}},
  \bibinfo {author} {\bibfnamefont {V.}~\bibnamefont {Umansky}}, \bibinfo
  {author} {\bibfnamefont {A.}~\bibnamefont {Stern}},\ and\ \bibinfo {author}
  {\bibfnamefont {D.}~\bibnamefont {Mahalu}},\ }\bibfield  {title} {\bibinfo
  {title} {Observation of a quarter of an electron charge at the nu = 5/2
  quantum hall state},\ }\href@noop {} {\bibfield  {journal} {\bibinfo
  {journal} {Nature}\ }\textbf {\bibinfo {volume} {452}},\ \bibinfo {pages}
  {829} (\bibinfo {year} {2008})}\BibitemShut {NoStop}%
\bibitem [{\citenamefont {Venkatachalam}\ \emph {et~al.}(2011)\citenamefont
  {Venkatachalam}, \citenamefont {Yacoby}, \citenamefont {Pfeiffer},\ and\
  \citenamefont {West}}]{venkatachalam2011local}%
  \BibitemOpen
  \bibfield  {author} {\bibinfo {author} {\bibfnamefont {V.}~\bibnamefont
  {Venkatachalam}}, \bibinfo {author} {\bibfnamefont {A.}~\bibnamefont
  {Yacoby}}, \bibinfo {author} {\bibfnamefont {L.}~\bibnamefont {Pfeiffer}},\
  and\ \bibinfo {author} {\bibfnamefont {K.}~\bibnamefont {West}},\ }\bibfield
  {title} {\bibinfo {title} {Local charge of the nu = 5/2 fractional quantum
  hall state},\ }\href@noop {} {\bibfield  {journal} {\bibinfo  {journal}
  {Nature}\ }\textbf {\bibinfo {volume} {469}},\ \bibinfo {pages} {185}
  (\bibinfo {year} {2011})}\BibitemShut {NoStop}%
\bibitem [{\citenamefont {Kim}\ \emph {et~al.}(2026)\citenamefont {Kim},
  \citenamefont {Dev}, \citenamefont {Shaer}, \citenamefont {Kumar},
  \citenamefont {Ilin}, \citenamefont {Haug}, \citenamefont {Iskoz},
  \citenamefont {Watanabe}, \citenamefont {Taniguchi}, \citenamefont {Mross}
  \emph {et~al.}}]{kim2026aharonov}%
  \BibitemOpen
  \bibfield  {author} {\bibinfo {author} {\bibfnamefont {J.}~\bibnamefont
  {Kim}}, \bibinfo {author} {\bibfnamefont {H.}~\bibnamefont {Dev}}, \bibinfo
  {author} {\bibfnamefont {A.}~\bibnamefont {Shaer}}, \bibinfo {author}
  {\bibfnamefont {R.}~\bibnamefont {Kumar}}, \bibinfo {author} {\bibfnamefont
  {A.}~\bibnamefont {Ilin}}, \bibinfo {author} {\bibfnamefont {A.}~\bibnamefont
  {Haug}}, \bibinfo {author} {\bibfnamefont {S.}~\bibnamefont {Iskoz}},
  \bibinfo {author} {\bibfnamefont {K.}~\bibnamefont {Watanabe}}, \bibinfo
  {author} {\bibfnamefont {T.}~\bibnamefont {Taniguchi}}, \bibinfo {author}
  {\bibfnamefont {D.~F.}\ \bibnamefont {Mross}}, \emph {et~al.},\ }\bibfield
  {title} {\bibinfo {title} {Aharonov--bohm interference in even-denominator
  fractional quantum hall states},\ }\href@noop {} {\bibfield  {journal}
  {\bibinfo  {journal} {Nature}\ }\textbf {\bibinfo {volume} {649}},\ \bibinfo
  {pages} {323} (\bibinfo {year} {2026})}\BibitemShut {NoStop}%
\bibitem [{\citenamefont {Read}\ and\ \citenamefont
  {Rezayi}(1999)}]{read1999beyond}%
  \BibitemOpen
  \bibfield  {author} {\bibinfo {author} {\bibfnamefont {N.}~\bibnamefont
  {Read}}\ and\ \bibinfo {author} {\bibfnamefont {E.}~\bibnamefont {Rezayi}},\
  }\bibfield  {title} {\bibinfo {title} {Beyond paired quantum hall states:
  Parafermions and incompressible states in the first excited landau level},\
  }\href@noop {} {\bibfield  {journal} {\bibinfo  {journal} {Physical Review
  B}\ }\textbf {\bibinfo {volume} {59}},\ \bibinfo {pages} {8084} (\bibinfo
  {year} {1999})}\BibitemShut {NoStop}%
\bibitem [{\citenamefont {Rezayi}\ and\ \citenamefont
  {Read}(2009)}]{rezayi2009non}%
  \BibitemOpen
  \bibfield  {author} {\bibinfo {author} {\bibfnamefont {E.}~\bibnamefont
  {Rezayi}}\ and\ \bibinfo {author} {\bibfnamefont {N.}~\bibnamefont {Read}},\
  }\bibfield  {title} {\bibinfo {title} {Non-abelian quantized hall states of
  electrons at filling factors 12/5 and 13/5 in the first excited landau
  level},\ }\href@noop {} {\bibfield  {journal} {\bibinfo  {journal} {Physical
  Review B—Condensed Matter and Materials Physics}\ }\textbf {\bibinfo
  {volume} {79}},\ \bibinfo {pages} {075306} (\bibinfo {year}
  {2009})}\BibitemShut {NoStop}%
\bibitem [{\citenamefont {W{\'o}js}(2009)}]{wojs2009transition}%
  \BibitemOpen
  \bibfield  {author} {\bibinfo {author} {\bibfnamefont {A.}~\bibnamefont
  {W{\'o}js}},\ }\bibfield  {title} {\bibinfo {title} {Transition from abelian
  to non-abelian quantum liquids in the second landau level},\ }\href@noop {}
  {\bibfield  {journal} {\bibinfo  {journal} {Physical Review B—Condensed
  Matter and Materials Physics}\ }\textbf {\bibinfo {volume} {80}},\ \bibinfo
  {pages} {041104} (\bibinfo {year} {2009})}\BibitemShut {NoStop}%
\bibitem [{\citenamefont {Zhu}\ \emph {et~al.}(2015)\citenamefont {Zhu},
  \citenamefont {Gong}, \citenamefont {Haldane},\ and\ \citenamefont
  {Sheng}}]{zhu2015fractional}%
  \BibitemOpen
  \bibfield  {author} {\bibinfo {author} {\bibfnamefont {W.}~\bibnamefont
  {Zhu}}, \bibinfo {author} {\bibfnamefont {S.}~\bibnamefont {Gong}}, \bibinfo
  {author} {\bibfnamefont {F.}~\bibnamefont {Haldane}},\ and\ \bibinfo {author}
  {\bibfnamefont {D.}~\bibnamefont {Sheng}},\ }\bibfield  {title} {\bibinfo
  {title} {Fractional quantum hall states at nu= 13/5 and 12/5 and their
  non-abelian nature},\ }\href@noop {} {\bibfield  {journal} {\bibinfo
  {journal} {Physical review letters}\ }\textbf {\bibinfo {volume} {115}},\
  \bibinfo {pages} {126805} (\bibinfo {year} {2015})}\BibitemShut {NoStop}%
\bibitem [{\citenamefont {Wu}\ \emph {et~al.}(2017)\citenamefont {Wu},
  \citenamefont {Shi},\ and\ \citenamefont {Jain}}]{Wu2017a}%
  \BibitemOpen
  \bibfield  {author} {\bibinfo {author} {\bibfnamefont {Y.~H.}\ \bibnamefont
  {Wu}}, \bibinfo {author} {\bibfnamefont {T.}~\bibnamefont {Shi}},\ and\
  \bibinfo {author} {\bibfnamefont {J.~K.}\ \bibnamefont {Jain}},\ }\bibfield
  {title} {\bibinfo {title} {Non-abelian parton fractional quantum hall effect
  in multilayer graphene},\ }\href
  {https://doi.org/10.1021/acs.nanolett.7b01080} {\bibfield  {journal}
  {\bibinfo  {journal} {Nano Letters}\ }\textbf {\bibinfo {volume} {17}},\
  \bibinfo {pages} {4643} (\bibinfo {year} {2017})}\BibitemShut {NoStop}%
\bibitem [{\citenamefont {Wang}\ and\ \citenamefont
  {Yang}(2022)}]{wang2022analytic}%
  \BibitemOpen
  \bibfield  {author} {\bibinfo {author} {\bibfnamefont {Y.}~\bibnamefont
  {Wang}}\ and\ \bibinfo {author} {\bibfnamefont {B.}~\bibnamefont {Yang}},\
  }\bibfield  {title} {\bibinfo {title} {Analytic exposition of the graviton
  modes in fractional quantum hall effects and its physical implications},\
  }\href@noop {} {\bibfield  {journal} {\bibinfo  {journal} {Physical Review
  B}\ }\textbf {\bibinfo {volume} {105}},\ \bibinfo {pages} {035144} (\bibinfo
  {year} {2022})}\BibitemShut {NoStop}%
\bibitem [{\citenamefont {Yang}(2022)}]{yang2022anyons}%
  \BibitemOpen
  \bibfield  {author} {\bibinfo {author} {\bibfnamefont {B.}~\bibnamefont
  {Yang}},\ }\bibfield  {title} {\bibinfo {title} {Anyons in conformal hilbert
  spaces: Statistics and dynamics of gapless excitations in fractional quantum
  hall systems},\ }\href@noop {} {\bibfield  {journal} {\bibinfo  {journal}
  {International Journal of Modern Physics B}\ }\textbf {\bibinfo {volume}
  {36}},\ \bibinfo {pages} {2230003} (\bibinfo {year} {2022})}\BibitemShut
  {NoStop}%
\bibitem [{\citenamefont {Laughlin}(1983)}]{laughlin1983anomalous}%
  \BibitemOpen
  \bibfield  {author} {\bibinfo {author} {\bibfnamefont {R.~B.}\ \bibnamefont
  {Laughlin}},\ }\bibfield  {title} {\bibinfo {title} {Anomalous quantum hall
  effect: An incompressible quantum fluid with fractionally charged
  excitations},\ }\href@noop {} {\bibfield  {journal} {\bibinfo  {journal}
  {Physical Review Letters}\ }\textbf {\bibinfo {volume} {50}},\ \bibinfo
  {pages} {1395} (\bibinfo {year} {1983})}\BibitemShut {NoStop}%
\bibitem [{\citenamefont {Cage}\ \emph {et~al.}(2012)\citenamefont {Cage},
  \citenamefont {Klitzing}, \citenamefont {Chang}, \citenamefont {Duncan},
  \citenamefont {Haldane}, \citenamefont {Laughlin}, \citenamefont {Pruisken},\
  and\ \citenamefont {Thouless}}]{cage2012quantum}%
  \BibitemOpen
  \bibfield  {author} {\bibinfo {author} {\bibfnamefont {M.~E.}\ \bibnamefont
  {Cage}}, \bibinfo {author} {\bibfnamefont {K.}~\bibnamefont {Klitzing}},
  \bibinfo {author} {\bibfnamefont {A.}~\bibnamefont {Chang}}, \bibinfo
  {author} {\bibfnamefont {F.}~\bibnamefont {Duncan}}, \bibinfo {author}
  {\bibfnamefont {M.}~\bibnamefont {Haldane}}, \bibinfo {author} {\bibfnamefont
  {R.~B.}\ \bibnamefont {Laughlin}}, \bibinfo {author} {\bibfnamefont
  {A.}~\bibnamefont {Pruisken}},\ and\ \bibinfo {author} {\bibfnamefont
  {D.}~\bibnamefont {Thouless}},\ }\href@noop {} {\emph {\bibinfo {title} {The
  quantum Hall effect}}}\ (\bibinfo  {publisher} {Springer Science \& Business
  Media},\ \bibinfo {year} {2012})\BibitemShut {NoStop}%
\bibitem [{\citenamefont {Haldane}(1983)}]{haldane1983fractional}%
  \BibitemOpen
  \bibfield  {author} {\bibinfo {author} {\bibfnamefont {F.~D.~M.}\
  \bibnamefont {Haldane}},\ }\bibfield  {title} {\bibinfo {title} {Fractional
  quantization of the hall effect: A hierarchy of incompressible quantum fluid
  states},\ }\href@noop {} {\bibfield  {journal} {\bibinfo  {journal} {Physical
  Review Letters}\ }\textbf {\bibinfo {volume} {51}},\ \bibinfo {pages} {605}
  (\bibinfo {year} {1983})}\BibitemShut {NoStop}%
\bibitem [{\citenamefont {Trung}\ and\ \citenamefont
  {Yang}(2021)}]{trung2021fractionalization}%
  \BibitemOpen
  \bibfield  {author} {\bibinfo {author} {\bibfnamefont {H.~Q.}\ \bibnamefont
  {Trung}}\ and\ \bibinfo {author} {\bibfnamefont {B.}~\bibnamefont {Yang}},\
  }\bibfield  {title} {\bibinfo {title} {Fractionalization and dynamics of
  anyons and their experimental signatures in the nu= n+ 1/3 fractional quantum
  hall state},\ }\href@noop {} {\bibfield  {journal} {\bibinfo  {journal}
  {Physical Review Letters}\ }\textbf {\bibinfo {volume} {127}},\ \bibinfo
  {pages} {046402} (\bibinfo {year} {2021})}\BibitemShut {NoStop}%
\bibitem [{\citenamefont {Wang}\ and\ \citenamefont
  {Yang}(2026)}]{wang2026microscopic}%
  \BibitemOpen
  \bibfield  {author} {\bibinfo {author} {\bibfnamefont {Y.}~\bibnamefont
  {Wang}}\ and\ \bibinfo {author} {\bibfnamefont {B.}~\bibnamefont {Yang}},\
  }\bibfield  {title} {\bibinfo {title} {Microscopic geometric theory for
  gapped excitations in fractional topological fluids},\ }\href@noop {}
  {\bibfield  {journal} {\bibinfo  {journal} {arXiv preprint arXiv:2603.13489}\
  } (\bibinfo {year} {2026})}\BibitemShut {NoStop}%
\bibitem [{\citenamefont {Balram}\ \emph {et~al.}(2020)\citenamefont {Balram},
  \citenamefont {Jain},\ and\ \citenamefont {Barkeshli}}]{balram2020z}%
  \BibitemOpen
  \bibfield  {author} {\bibinfo {author} {\bibfnamefont {A.~C.}\ \bibnamefont
  {Balram}}, \bibinfo {author} {\bibfnamefont {J.}~\bibnamefont {Jain}},\ and\
  \bibinfo {author} {\bibfnamefont {M.}~\bibnamefont {Barkeshli}},\ }\bibfield
  {title} {\bibinfo {title} {Z n superconductivity of composite bosons and the
  7/3 fractional quantum hall effect},\ }\href@noop {} {\bibfield  {journal}
  {\bibinfo  {journal} {Physical Review Research}\ }\textbf {\bibinfo {volume}
  {2}},\ \bibinfo {pages} {013349} (\bibinfo {year} {2020})}\BibitemShut
  {NoStop}%
\bibitem [{\citenamefont {Simon}\ \emph {et~al.}(2007)\citenamefont {Simon},
  \citenamefont {Rezayi},\ and\ \citenamefont {Cooper}}]{simon2007generalized}%
  \BibitemOpen
  \bibfield  {author} {\bibinfo {author} {\bibfnamefont {S.~H.}\ \bibnamefont
  {Simon}}, \bibinfo {author} {\bibfnamefont {E.}~\bibnamefont {Rezayi}},\ and\
  \bibinfo {author} {\bibfnamefont {N.~R.}\ \bibnamefont {Cooper}},\ }\bibfield
   {title} {\bibinfo {title} {Generalized quantum hall projection
  hamiltonians},\ }\href@noop {} {\bibfield  {journal} {\bibinfo  {journal}
  {Physical Review B—Condensed Matter and Materials Physics}\ }\textbf
  {\bibinfo {volume} {75}},\ \bibinfo {pages} {075318} (\bibinfo {year}
  {2007})}\BibitemShut {NoStop}%
\bibitem [{\citenamefont {{See supplementary material for detailed calculation
  and analysis}}()}]{seesup}%
  \BibitemOpen
  \bibfield  {author} {\bibinfo {author} {\bibnamefont {{See supplementary
  material for detailed calculation and analysis}}},\ }\href@noop {}
  {}\BibitemShut {NoStop}%
\bibitem [{\citenamefont {Xu}\ \emph {et~al.}(2025)\citenamefont {Xu},
  \citenamefont {Ji}, \citenamefont {Wang}, \citenamefont {Trung},\ and\
  \citenamefont {Yang}}]{xu2025dynamics}%
  \BibitemOpen
  \bibfield  {author} {\bibinfo {author} {\bibfnamefont {Q.}~\bibnamefont
  {Xu}}, \bibinfo {author} {\bibfnamefont {G.}~\bibnamefont {Ji}}, \bibinfo
  {author} {\bibfnamefont {Y.}~\bibnamefont {Wang}}, \bibinfo {author}
  {\bibfnamefont {H.~Q.}\ \bibnamefont {Trung}},\ and\ \bibinfo {author}
  {\bibfnamefont {B.}~\bibnamefont {Yang}},\ }\bibfield  {title} {\bibinfo
  {title} {Dynamics of anyon clusters in fractional quantum hall fluids},\
  }\href@noop {} {\bibfield  {journal} {\bibinfo  {journal} {Physical Review
  B}\ }\textbf {\bibinfo {volume} {112}},\ \bibinfo {pages} {235112} (\bibinfo
  {year} {2025})}\BibitemShut {NoStop}%
\bibitem [{\citenamefont {Cooper}(1956)}]{cooper1956bound}%
  \BibitemOpen
  \bibfield  {author} {\bibinfo {author} {\bibfnamefont {L.~N.}\ \bibnamefont
  {Cooper}},\ }\bibfield  {title} {\bibinfo {title} {Bound electron pairs in a
  degenerate fermi gas},\ }\href@noop {} {\bibfield  {journal} {\bibinfo
  {journal} {Physical Review}\ }\textbf {\bibinfo {volume} {104}},\ \bibinfo
  {pages} {1189} (\bibinfo {year} {1956})}\BibitemShut {NoStop}%
\bibitem [{\citenamefont {Bardeen}\ \emph {et~al.}(1957)\citenamefont
  {Bardeen}, \citenamefont {Cooper},\ and\ \citenamefont
  {Schrieffer}}]{bardeen1957microscopic}%
  \BibitemOpen
  \bibfield  {author} {\bibinfo {author} {\bibfnamefont {J.}~\bibnamefont
  {Bardeen}}, \bibinfo {author} {\bibfnamefont {L.~N.}\ \bibnamefont
  {Cooper}},\ and\ \bibinfo {author} {\bibfnamefont {J.~R.}\ \bibnamefont
  {Schrieffer}},\ }\bibfield  {title} {\bibinfo {title} {Microscopic theory of
  superconductivity},\ }\href@noop {} {\bibfield  {journal} {\bibinfo
  {journal} {Physical Review}\ }\textbf {\bibinfo {volume} {106}},\ \bibinfo
  {pages} {162} (\bibinfo {year} {1957})}\BibitemShut {NoStop}%
\bibitem [{\citenamefont {Yang}(2013)}]{yang2013analytic}%
  \BibitemOpen
  \bibfield  {author} {\bibinfo {author} {\bibfnamefont {B.}~\bibnamefont
  {Yang}},\ }\bibfield  {title} {\bibinfo {title} {Analytic wave functions for
  neutral bulk excitations in fractional quantum hall fluids},\ }\href@noop {}
  {\bibfield  {journal} {\bibinfo  {journal} {Physical Review B—Condensed
  Matter and Materials Physics}\ }\textbf {\bibinfo {volume} {87}},\ \bibinfo
  {pages} {245132} (\bibinfo {year} {2013})}\BibitemShut {NoStop}%
\bibitem [{\citenamefont {Li}\ and\ \citenamefont
  {Haldane}(2008)}]{li2008entanglement}%
  \BibitemOpen
  \bibfield  {author} {\bibinfo {author} {\bibfnamefont {H.}~\bibnamefont
  {Li}}\ and\ \bibinfo {author} {\bibfnamefont {F.~D.~M.}\ \bibnamefont
  {Haldane}},\ }\bibfield  {title} {\bibinfo {title} {Entanglement spectrum as
  a generalization of entanglement entropy: Identification of topological order
  in non-abelian fractional quantum hall effect states},\ }\href@noop {}
  {\bibfield  {journal} {\bibinfo  {journal} {Physical review letters}\
  }\textbf {\bibinfo {volume} {101}},\ \bibinfo {pages} {010504} (\bibinfo
  {year} {2008})}\BibitemShut {NoStop}%
\bibitem [{\citenamefont {Sohal}\ \emph {et~al.}(2020)\citenamefont {Sohal},
  \citenamefont {Han}, \citenamefont {Santos},\ and\ \citenamefont
  {Teo}}]{sohal2020entanglement}%
  \BibitemOpen
  \bibfield  {author} {\bibinfo {author} {\bibfnamefont {R.}~\bibnamefont
  {Sohal}}, \bibinfo {author} {\bibfnamefont {B.}~\bibnamefont {Han}}, \bibinfo
  {author} {\bibfnamefont {L.~H.}\ \bibnamefont {Santos}},\ and\ \bibinfo
  {author} {\bibfnamefont {J.~C.}\ \bibnamefont {Teo}},\ }\bibfield  {title}
  {\bibinfo {title} {Entanglement entropy of generalized moore-read fractional
  quantum hall state interfaces},\ }\href@noop {} {\bibfield  {journal}
  {\bibinfo  {journal} {Physical Review B}\ }\textbf {\bibinfo {volume}
  {102}},\ \bibinfo {pages} {045102} (\bibinfo {year} {2020})}\BibitemShut
  {NoStop}%
\bibitem [{\citenamefont {Halperin}(1983)}]{halperin1983theory}%
  \BibitemOpen
  \bibfield  {author} {\bibinfo {author} {\bibfnamefont {B.~I.}\ \bibnamefont
  {Halperin}},\ }\bibfield  {title} {\bibinfo {title} {Theory of the quantized
  hall conductance},\ }\href@noop {} {\bibfield  {journal} {\bibinfo  {journal}
  {helv. phys. acta}\ }\textbf {\bibinfo {volume} {56}},\ \bibinfo {pages} {75}
  (\bibinfo {year} {1983})}\BibitemShut {NoStop}%
\bibitem [{\citenamefont {Lopez}\ and\ \citenamefont
  {Fradkin}(2001)}]{lopez2001effective}%
  \BibitemOpen
  \bibfield  {author} {\bibinfo {author} {\bibfnamefont {A.}~\bibnamefont
  {Lopez}}\ and\ \bibinfo {author} {\bibfnamefont {E.}~\bibnamefont
  {Fradkin}},\ }\bibfield  {title} {\bibinfo {title} {Effective field theory
  for the bulk and edge states of quantum hall states in unpolarized single
  layer and bilayer systems},\ }\href@noop {} {\bibfield  {journal} {\bibinfo
  {journal} {Physical Review B}\ }\textbf {\bibinfo {volume} {63}},\ \bibinfo
  {pages} {085306} (\bibinfo {year} {2001})}\BibitemShut {NoStop}%
\bibitem [{\citenamefont {Trung}\ \emph {et~al.}(2025)\citenamefont {Trung},
  \citenamefont {Xu},\ and\ \citenamefont {Yang}}]{trung2025long}%
  \BibitemOpen
  \bibfield  {author} {\bibinfo {author} {\bibfnamefont {H.~Q.}\ \bibnamefont
  {Trung}}, \bibinfo {author} {\bibfnamefont {Q.}~\bibnamefont {Xu}},\ and\
  \bibinfo {author} {\bibfnamefont {B.}~\bibnamefont {Yang}},\ }\bibfield
  {title} {\bibinfo {title} {Long-range entanglement and role of realistic
  interaction in braiding of non-abelian quasiholes in fractional quantum hall
  phases},\ }\href {https://doi.org/10.1103/c4ck-8rcx} {\bibfield  {journal}
  {\bibinfo  {journal} {Phys. Rev. B}\ }\textbf {\bibinfo {volume} {112}},\
  \bibinfo {pages} {205101} (\bibinfo {year} {2025})}\BibitemShut {NoStop}%
\bibitem [{\citenamefont {Storni}\ and\ \citenamefont
  {Morf}(2011)}]{storni2011localized}%
  \BibitemOpen
  \bibfield  {author} {\bibinfo {author} {\bibfnamefont {M.}~\bibnamefont
  {Storni}}\ and\ \bibinfo {author} {\bibfnamefont {R.}~\bibnamefont {Morf}},\
  }\bibfield  {title} {\bibinfo {title} {Localized quasiholes and the majorana
  fermion in fractional quantum hall state at nu= 5 2 via direct
  diagonalization},\ }\href@noop {} {\bibfield  {journal} {\bibinfo  {journal}
  {Physical Review B}\ }\textbf {\bibinfo {volume} {83}},\ \bibinfo {pages}
  {195306} (\bibinfo {year} {2011})}\BibitemShut {NoStop}%
\bibitem [{\citenamefont {Greiter}(2011)}]{greiter2011landau}%
  \BibitemOpen
  \bibfield  {author} {\bibinfo {author} {\bibfnamefont {M.}~\bibnamefont
  {Greiter}},\ }\bibfield  {title} {\bibinfo {title} {Landau level quantization
  on the sphere},\ }\href@noop {} {\bibfield  {journal} {\bibinfo  {journal}
  {Physical Review B}\ }\textbf {\bibinfo {volume} {83}},\ \bibinfo {pages}
  {115129} (\bibinfo {year} {2011})}\BibitemShut {NoStop}%
\bibitem [{\citenamefont {Yang}(2021)}]{yang2021statistical}%
  \BibitemOpen
  \bibfield  {author} {\bibinfo {author} {\bibfnamefont {B.}~\bibnamefont
  {Yang}},\ }\bibfield  {title} {\bibinfo {title} {Statistical interactions and
  boson-anyon duality in fractional quantum hall fluids},\ }\href@noop {}
  {\bibfield  {journal} {\bibinfo  {journal} {Physical Review Letters}\
  }\textbf {\bibinfo {volume} {127}},\ \bibinfo {pages} {126406} (\bibinfo
  {year} {2021})}\BibitemShut {NoStop}%
\bibitem [{\citenamefont {Wu}\ \emph {et~al.}(2014)\citenamefont {Wu},
  \citenamefont {Estienne}, \citenamefont {Regnault},\ and\ \citenamefont
  {Bernevig}}]{wu2014braiding}%
  \BibitemOpen
  \bibfield  {author} {\bibinfo {author} {\bibfnamefont {Y.-L.}\ \bibnamefont
  {Wu}}, \bibinfo {author} {\bibfnamefont {B.}~\bibnamefont {Estienne}},
  \bibinfo {author} {\bibfnamefont {N.}~\bibnamefont {Regnault}},\ and\
  \bibinfo {author} {\bibfnamefont {B.~A.}\ \bibnamefont {Bernevig}},\
  }\bibfield  {title} {\bibinfo {title} {Braiding non-abelian quasiholes in
  fractional quantum hall states},\ }\href@noop {} {\bibfield  {journal}
  {\bibinfo  {journal} {Physical review letters}\ }\textbf {\bibinfo {volume}
  {113}},\ \bibinfo {pages} {116801} (\bibinfo {year} {2014})}\BibitemShut
  {NoStop}%
\bibitem [{\citenamefont {Johri}\ \emph {et~al.}(2014)\citenamefont {Johri},
  \citenamefont {Papi{\'c}}, \citenamefont {Bhatt},\ and\ \citenamefont
  {Schmitteckert}}]{johri2014quasiholes}%
  \BibitemOpen
  \bibfield  {author} {\bibinfo {author} {\bibfnamefont {S.}~\bibnamefont
  {Johri}}, \bibinfo {author} {\bibfnamefont {Z.}~\bibnamefont {Papi{\'c}}},
  \bibinfo {author} {\bibfnamefont {R.~N.}\ \bibnamefont {Bhatt}},\ and\
  \bibinfo {author} {\bibfnamefont {P.}~\bibnamefont {Schmitteckert}},\
  }\bibfield  {title} {\bibinfo {title} {Quasiholes of 1 3 and 7 3 quantum hall
  states: Size estimates via exact diagonalization and density-matrix
  renormalization group},\ }\href@noop {} {\bibfield  {journal} {\bibinfo
  {journal} {Physical Review B}\ }\textbf {\bibinfo {volume} {89}},\ \bibinfo
  {pages} {115124} (\bibinfo {year} {2014})}\BibitemShut {NoStop}%
\bibitem [{\citenamefont {Li}\ \emph {et~al.}(2022)\citenamefont {Li},
  \citenamefont {Ye}, \citenamefont {Jiang}, \citenamefont {Jiang},
  \citenamefont {Wan},\ and\ \citenamefont {Hu}}]{li2022anyonic}%
  \BibitemOpen
  \bibfield  {author} {\bibinfo {author} {\bibfnamefont {J.}~\bibnamefont
  {Li}}, \bibinfo {author} {\bibfnamefont {D.}~\bibnamefont {Ye}}, \bibinfo
  {author} {\bibfnamefont {C.-X.}\ \bibnamefont {Jiang}}, \bibinfo {author}
  {\bibfnamefont {N.}~\bibnamefont {Jiang}}, \bibinfo {author} {\bibfnamefont
  {X.}~\bibnamefont {Wan}},\ and\ \bibinfo {author} {\bibfnamefont {Z.-X.}\
  \bibnamefont {Hu}},\ }\bibfield  {title} {\bibinfo {title} {Anyonic braiding
  via quench dynamics in fractional quantum hall liquids},\ }\href@noop {}
  {\bibfield  {journal} {\bibinfo  {journal} {Physical Review B}\ }\textbf
  {\bibinfo {volume} {105}},\ \bibinfo {pages} {195311} (\bibinfo {year}
  {2022})}\BibitemShut {NoStop}%
\bibitem [{\citenamefont {Umucalllar}\ \emph {et~al.}(2018)\citenamefont
  {Umucalllar}, \citenamefont {Macaluso}, \citenamefont {Comparin},\ and\
  \citenamefont {Carusotto}}]{Umucalllar2018}%
  \BibitemOpen
  \bibfield  {author} {\bibinfo {author} {\bibfnamefont {R.~O.}\ \bibnamefont
  {Umucalllar}}, \bibinfo {author} {\bibfnamefont {E.}~\bibnamefont
  {Macaluso}}, \bibinfo {author} {\bibfnamefont {T.}~\bibnamefont {Comparin}},\
  and\ \bibinfo {author} {\bibfnamefont {I.}~\bibnamefont {Carusotto}},\
  }\bibfield  {title} {\bibinfo {title} {Time-of-flight measurements as a
  possible method to observe anyonic statistics},\ }\bibfield  {journal}
  {\bibinfo  {journal} {Physical Review Letters}\ }\textbf {\bibinfo {volume}
  {120}},\ \href {https://doi.org/10.1103/PhysRevLett.120.230403}
  {10.1103/PhysRevLett.120.230403} (\bibinfo {year} {2018})\BibitemShut
  {NoStop}%
\bibitem [{\citenamefont {Macaluso}\ \emph {et~al.}(2019)\citenamefont
  {Macaluso}, \citenamefont {Comparin}, \citenamefont {Mazza},\ and\
  \citenamefont {Carusotto}}]{Macaluso2019}%
  \BibitemOpen
  \bibfield  {author} {\bibinfo {author} {\bibfnamefont {E.}~\bibnamefont
  {Macaluso}}, \bibinfo {author} {\bibfnamefont {T.}~\bibnamefont {Comparin}},
  \bibinfo {author} {\bibfnamefont {L.}~\bibnamefont {Mazza}},\ and\ \bibinfo
  {author} {\bibfnamefont {I.}~\bibnamefont {Carusotto}},\ }\bibfield  {title}
  {\bibinfo {title} {Fusion channels of non-abelian anyons from
  angular-momentum and density-profile measurements},\ }\bibfield  {journal}
  {\bibinfo  {journal} {Physical Review Letters}\ }\textbf {\bibinfo {volume}
  {123}},\ \href {https://doi.org/10.1103/PhysRevLett.123.266801}
  {10.1103/PhysRevLett.123.266801} (\bibinfo {year} {2019})\BibitemShut
  {NoStop}%
\bibitem [{\citenamefont {Comparin}\ \emph {et~al.}(2021)\citenamefont
  {Comparin}, \citenamefont {Opler}, \citenamefont {Macaluso}, \citenamefont
  {Biella}, \citenamefont {Polychronakos},\ and\ \citenamefont
  {Mazza}}]{Comparin2021}%
  \BibitemOpen
  \bibfield  {author} {\bibinfo {author} {\bibfnamefont {T.}~\bibnamefont
  {Comparin}}, \bibinfo {author} {\bibfnamefont {A.}~\bibnamefont {Opler}},
  \bibinfo {author} {\bibfnamefont {E.}~\bibnamefont {Macaluso}}, \bibinfo
  {author} {\bibfnamefont {A.}~\bibnamefont {Biella}}, \bibinfo {author}
  {\bibfnamefont {A.~P.}\ \bibnamefont {Polychronakos}},\ and\ \bibinfo
  {author} {\bibfnamefont {L.}~\bibnamefont {Mazza}},\ }\bibfield  {title}
  {\bibinfo {title} {A measurable fractional spin for quantum hall
  quasiparticles on the disk},\ }\href {http://arxiv.org/abs/2112.02901} {\ ,\
  \bibinfo {pages} {1} (\bibinfo {year} {2021})}\BibitemShut {NoStop}%
\bibitem [{\citenamefont {Trung}\ \emph {et~al.}(2023)\citenamefont {Trung},
  \citenamefont {Wang},\ and\ \citenamefont {Yang}}]{trung2023spin}%
  \BibitemOpen
  \bibfield  {author} {\bibinfo {author} {\bibfnamefont {H.~Q.}\ \bibnamefont
  {Trung}}, \bibinfo {author} {\bibfnamefont {Y.}~\bibnamefont {Wang}},\ and\
  \bibinfo {author} {\bibfnamefont {B.}~\bibnamefont {Yang}},\ }\bibfield
  {title} {\bibinfo {title} {Spin-statistics relation and abelian braiding
  phase for anyons in the fractional quantum hall effect},\ }\href@noop {}
  {\bibfield  {journal} {\bibinfo  {journal} {Physical Review B}\ }\textbf
  {\bibinfo {volume} {107}},\ \bibinfo {pages} {L201301} (\bibinfo {year}
  {2023})}\BibitemShut {NoStop}%
\bibitem [{\citenamefont {Ji}\ \emph {et~al.}(2024)\citenamefont {Ji},
  \citenamefont {Bose}, \citenamefont {Balram},\ and\ \citenamefont
  {Yang}}]{ji2024universal}%
  \BibitemOpen
  \bibfield  {author} {\bibinfo {author} {\bibfnamefont {G.}~\bibnamefont
  {Ji}}, \bibinfo {author} {\bibfnamefont {K.}~\bibnamefont {Bose}}, \bibinfo
  {author} {\bibfnamefont {A.~C.}\ \bibnamefont {Balram}},\ and\ \bibinfo
  {author} {\bibfnamefont {B.}~\bibnamefont {Yang}},\ }\bibfield  {title}
  {\bibinfo {title} {Universal modeling of oscillations in fractional quantum
  hall fluids},\ }\href@noop {} {\bibfield  {journal} {\bibinfo  {journal}
  {Physical Review B}\ }\textbf {\bibinfo {volume} {110}},\ \bibinfo {pages}
  {075113} (\bibinfo {year} {2024})}\BibitemShut {NoStop}%
\bibitem [{\citenamefont {Papi{\'c}}\ \emph {et~al.}(2018)\citenamefont
  {Papi{\'c}}, \citenamefont {Mong}, \citenamefont {Yazdani},\ and\
  \citenamefont {Zaletel}}]{papic2018imaging}%
  \BibitemOpen
  \bibfield  {author} {\bibinfo {author} {\bibfnamefont {Z.}~\bibnamefont
  {Papi{\'c}}}, \bibinfo {author} {\bibfnamefont {R.~S.}\ \bibnamefont {Mong}},
  \bibinfo {author} {\bibfnamefont {A.}~\bibnamefont {Yazdani}},\ and\ \bibinfo
  {author} {\bibfnamefont {M.~P.}\ \bibnamefont {Zaletel}},\ }\bibfield
  {title} {\bibinfo {title} {Imaging anyons with scanning tunneling
  microscopy},\ }\href@noop {} {\bibfield  {journal} {\bibinfo  {journal}
  {Physical Review X}\ }\textbf {\bibinfo {volume} {8}},\ \bibinfo {pages}
  {011037} (\bibinfo {year} {2018})}\BibitemShut {NoStop}%
\bibitem [{\citenamefont {Schilling}(2000)}]{schilling2000review}%
  \BibitemOpen
  \bibfield  {author} {\bibinfo {author} {\bibfnamefont {K.}~\bibnamefont
  {Schilling}},\ }\bibfield  {title} {\bibinfo {title} {Review on string
  breaking—the query in quest of the evidence},\ }\href@noop {} {\bibfield
  {journal} {\bibinfo  {journal} {Nuclear Physics B-Proceedings Supplements}\
  }\textbf {\bibinfo {volume} {83}},\ \bibinfo {pages} {140} (\bibinfo {year}
  {2000})}\BibitemShut {NoStop}%
\bibitem [{\citenamefont {Bali}\ \emph {et~al.}(2005)\citenamefont {Bali},
  \citenamefont {Neff}, \citenamefont {Duessel}, \citenamefont {Lippert},
  \citenamefont {Schilling},\ and\ \citenamefont
  {Collaboration)}}]{bali2005observation}%
  \BibitemOpen
  \bibfield  {author} {\bibinfo {author} {\bibfnamefont {G.~S.}\ \bibnamefont
  {Bali}}, \bibinfo {author} {\bibfnamefont {H.}~\bibnamefont {Neff}}, \bibinfo
  {author} {\bibfnamefont {T.}~\bibnamefont {Duessel}}, \bibinfo {author}
  {\bibfnamefont {T.}~\bibnamefont {Lippert}}, \bibinfo {author} {\bibfnamefont
  {K.}~\bibnamefont {Schilling}},\ and\ \bibinfo {author} {\bibfnamefont
  {S.}~\bibnamefont {Collaboration)}},\ }\bibfield  {title} {\bibinfo {title}
  {Observation of string breaking in qcd},\ }\href@noop {} {\bibfield
  {journal} {\bibinfo  {journal} {Physical Review D—Particles, Fields,
  Gravitation, and Cosmology}\ }\textbf {\bibinfo {volume} {71}},\ \bibinfo
  {pages} {114513} (\bibinfo {year} {2005})}\BibitemShut {NoStop}%
\bibitem [{\citenamefont {Jain}(1989)}]{jain1989incompressible}%
  \BibitemOpen
  \bibfield  {author} {\bibinfo {author} {\bibfnamefont {J.~K.}\ \bibnamefont
  {Jain}},\ }\bibfield  {title} {\bibinfo {title} {Incompressible quantum hall
  states},\ }\href@noop {} {\bibfield  {journal} {\bibinfo  {journal} {Physical
  Review B}\ }\textbf {\bibinfo {volume} {40}},\ \bibinfo {pages} {8079}
  (\bibinfo {year} {1989})}\BibitemShut {NoStop}%
\bibitem [{\citenamefont {Read}\ and\ \citenamefont {Moore}(1992)}]{Read1992}%
  \BibitemOpen
  \bibfield  {author} {\bibinfo {author} {\bibfnamefont {N.}~\bibnamefont
  {Read}}\ and\ \bibinfo {author} {\bibfnamefont {G.}~\bibnamefont {Moore}},\
  }\bibfield  {title} {\bibinfo {title} {Fractional quantum hall effect and
  nonabelian statistics},\ }\href {https://doi.org/10.1143/ptps.107.157}
  {\bibfield  {journal} {\bibinfo  {journal} {Progress of Theoretical Physics
  Supplement}\ }\textbf {\bibinfo {volume} {107}},\ \bibinfo {pages} {157}
  (\bibinfo {year} {1992})}\BibitemShut {NoStop}%
\bibitem [{\citenamefont {Bonderson}\ \emph {et~al.}(2011)\citenamefont
  {Bonderson}, \citenamefont {Gurarie},\ and\ \citenamefont
  {Nayak}}]{Bonderson2011}%
  \BibitemOpen
  \bibfield  {author} {\bibinfo {author} {\bibfnamefont {P.}~\bibnamefont
  {Bonderson}}, \bibinfo {author} {\bibfnamefont {V.}~\bibnamefont {Gurarie}},\
  and\ \bibinfo {author} {\bibfnamefont {C.}~\bibnamefont {Nayak}},\ }\bibfield
   {title} {\bibinfo {title} {Plasma analogy and non-abelian statistics for
  ising-type quantum hall states}\ }\textbf {\bibinfo {volume} {075303}},\
  \href {https://doi.org/10.1103/PhysRevB.83.075303}
  {10.1103/PhysRevB.83.075303} (\bibinfo {year} {2011})\BibitemShut {NoStop}%
\bibitem [{\citenamefont {Tserkovnyak}\ and\ \citenamefont
  {Simon}(2003)}]{Tserkovnyak2003}%
  \BibitemOpen
  \bibfield  {author} {\bibinfo {author} {\bibfnamefont {Y.}~\bibnamefont
  {Tserkovnyak}}\ and\ \bibinfo {author} {\bibfnamefont {S.~H.}\ \bibnamefont
  {Simon}},\ }\bibfield  {title} {\bibinfo {title} {Monte carlo evaluation of
  non-abelian statistics},\ }\href
  {https://doi.org/10.1103/PhysRevLett.90.016802} {\bibfield  {journal}
  {\bibinfo  {journal} {Physical Review Letters}\ }\textbf {\bibinfo {volume}
  {90}},\ \bibinfo {pages} {4} (\bibinfo {year} {2003})}\BibitemShut {NoStop}%
\bibitem [{\citenamefont {Bernevig}\ and\ \citenamefont
  {Haldane}(2008{\natexlab{a}})}]{bernevig2008model}%
  \BibitemOpen
  \bibfield  {author} {\bibinfo {author} {\bibfnamefont {B.~A.}\ \bibnamefont
  {Bernevig}}\ and\ \bibinfo {author} {\bibfnamefont {F.}~\bibnamefont
  {Haldane}},\ }\bibfield  {title} {\bibinfo {title} {Model fractional quantum
  hall states and jack polynomials},\ }\href@noop {} {\bibfield  {journal}
  {\bibinfo  {journal} {Physical review letters}\ }\textbf {\bibinfo {volume}
  {100}},\ \bibinfo {pages} {246802} (\bibinfo {year}
  {2008}{\natexlab{a}})}\BibitemShut {NoStop}%
\bibitem [{\citenamefont {Bernevig}\ and\ \citenamefont
  {Haldane}(2008{\natexlab{b}})}]{bernevig2008generalized}%
  \BibitemOpen
  \bibfield  {author} {\bibinfo {author} {\bibfnamefont {B.~A.}\ \bibnamefont
  {Bernevig}}\ and\ \bibinfo {author} {\bibfnamefont {F.}~\bibnamefont
  {Haldane}},\ }\bibfield  {title} {\bibinfo {title} {Generalized clustering
  conditions of jack polynomials at negative jack parameter $\alpha$},\
  }\href@noop {} {\bibfield  {journal} {\bibinfo  {journal} {Physical Review
  B}\ }\textbf {\bibinfo {volume} {77}},\ \bibinfo {pages} {184502} (\bibinfo
  {year} {2008}{\natexlab{b}})}\BibitemShut {NoStop}%
\bibitem [{Dia()}]{DiagHam}%
  \BibitemOpen
  \href {https://nick-ux.org/diagham} {\bibinfo {title} {Diagham}}\BibitemShut
  {NoStop}%
\bibitem [{\citenamefont {Rezayi}\ and\ \citenamefont
  {Haldane}(1985)}]{rezayi1985incompressible}%
  \BibitemOpen
  \bibfield  {author} {\bibinfo {author} {\bibfnamefont {E.}~\bibnamefont
  {Rezayi}}\ and\ \bibinfo {author} {\bibfnamefont {F.}~\bibnamefont
  {Haldane}},\ }\bibfield  {title} {\bibinfo {title} {Incompressible states of
  the fractionally quantized hall effect in the presence of impurities: A
  finite-size study},\ }\href@noop {} {\bibfield  {journal} {\bibinfo
  {journal} {Physical Review B}\ }\textbf {\bibinfo {volume} {32}},\ \bibinfo
  {pages} {6924} (\bibinfo {year} {1985})}\BibitemShut {NoStop}%
\bibitem [{\citenamefont {Green}(2001)}]{green2001strongly}%
  \BibitemOpen
  \bibfield  {author} {\bibinfo {author} {\bibfnamefont {D.}~\bibnamefont
  {Green}},\ }\href@noop {} {\emph {\bibinfo {title} {Strongly correlated
  states in low dimensions}}}\ (\bibinfo  {publisher} {Yale University},\
  \bibinfo {year} {2001})\BibitemShut {NoStop}%
\end{thebibliography}%

\end{document}